\documentclass[preprint, 10pt]{aastex}

\shorttitle{NGC~604}
\shortauthors{Bruhweiler et al.}

\begin{document}

\title{STIS Spectral Imagery of the OB Stars in NGC~604: The Most Luminous Stars}

\author{Fred C. Bruhweiler\altaffilmark{1}, Cherie L. Miskey\altaffilmark{1}, and Margaret Smith Neubig\altaffilmark{1}}
\affil{Institute for Astrophysics \& Computational Sciences}
\affil{Department of Physics, The Catholic University of America, Washington, DC 20064}
\email{bruhweiler@cua.edu, miskey@iacs.gsfc.nasa.gov, neubig@stars.gsfc.nasa.gov}


\altaffiltext{1}{Laboratory for Astronomy and Solar Physics, 
NASA/Goddard Space Flight Center, Greenbelt, MD 20771}

\begin{abstract}

We present results using two-dimensional spectral imagery and photometry obtained with the Hubble Space Telescope (HST) for the starburst H~II region, NGC~604, in the nearby galaxy M33. The spectral imagery was acquired with the Space Telescope Imaging Spectrograph (STIS) using the MAMA/G140L configuration, which provided wavelength coverage spanning 1170-1730\AA. From a single 1720 second STIS exposure, we have extracted spectra for 49 stars and derived individual UV spectral types for 40 stars in the crowded 25$\arcsec$ x 2$\arcsec$ stellar field sampled by the STIS aperture. These stars represent a significant fraction of the young, luminous O and B stars in NGC~604. Three objects have pronounced He~II $\lambda$1640 emission, the signature of W-R or luminous Of stars. By combining UV fluxes with HST WFPC and WFPC2 photometry at visual wavelengths, we derive the extinction curve for NGC~604. We use this extinction curve, together with the available accurate distance for M33, derived UV spectral types, and HST photometry, to determine positions of the luminous stars in the upper H-R diagram for NGC~604. The revision to the O star effective temperature scale by Martins et al., based upon non-LTE, line-blanketed, model atmospheres, is essential in obtaining reliable positions in the log(L$_{*}$)-log(T$_{\rm eff}$) plane.  These stars are quite young with a characteristic age of $\sim$ 3 Myr. The spectra and photometry indicate three objects are exceedingly luminous. Their inferred locations in the H-R diagram relative to theoretical evolutionary tracks indicate stellar masses $\geq$ 120 M$_{\sun}$. High spatial resolution HST imagery provides no evidence of multiple stars composing these objects. Still we cannot eliminate the possibility that these objects are not unresolved multiple stars of lower mass, possibly W-R stars. Simple tests demonstrate that the ten most luminous stars predominantly determine the UV spectral features seen in the total light of NGC~604. We conclude that the interpretation of spectral fitting of more distant starburst galaxies, where individual stars are not resolved, must be done with extreme care. 
\end{abstract}





\section{INTRODUCTION}
Recent theoretical studies maintain that the first generation Population III stars were quite massive, possibly 100-1000 M$_{\sun}$ (cf. Schaerer et al. 2002; Nakamura \& Umemura 2001; Ostriker \& Gnedin 1996). This has rekindled interest in finding strong observational constraints for the formation of the most massive stars. Observations of individual massive stars in nearby giant star forming regions such as 30 Dor and NGC~604, even in the present-day higher metallicity Universe, are extremely important. Such observations potentially can provide two critical observational parameters; the shape of the Initial Mass Function (IMF) for massive stars and the upper mass cutoff for the main sequence. 

The bright H II region, NGC~604, in the nearby spiral galaxy M 33 (d= 840 Kpc; 
Freedman, Wilson, \& Madore 1991), and 30 Doradus, 
in the Large Magellanic Cloud (LMC), are the two most luminous concentrations 
of OB stars in the Local Group. Their luminosities are comparable to many starbursts in more distant galaxies outside the Local Group. Moreover, both regions have metallicities that are sub-solar, with roughly 0.25 and 0.4 Z$_{\sun}$ for the LMC (Dufour, Shields, \& Talbot 1985) and NGC~604 (Vilchez et al. 1985), respectively. These low metallicities may make it possible to use 30 Dor and NGC~604 as surrogate laboratories for understanding the properties of starburst activity seen at higher redshift in the Early Universe. 

Both 30 Doradus and NGC~604 are quite luminous, but they differ in that stars of 30 Dor are mostly concentrated in a tight, compact cluster, while stars in the more distant NGC~604 of M33 are loosely dispersed. Consequently, even though NGC~604 is more distant than 30 Dor in the LMC (840 Kpc versus 52 Kpc), individual stars in both star forming regions can be spatially resolved and studied using the high angular resolution and two-dimensional spectral imaging capabilities of the STIS aboard the HST. In a companion paper (Miskey \& Bruhweiler 2003; Hereafter Paper 1), we describe the software and algorithms that we have developed to separate and disentangle overlapping spectra with angular separations as small as 0.05 $\arcsec$. Using the newly developed software and the capabilities of the STIS, we can now study efficiently a large number of individual stars in star-forming regions throughout the Local Group.

We have embarked upon a detailed study utilizing the STIS to probe the central regions of both NGC~604 and 30 Dor as well as other OB stellar concentrations in the Local Group. In this paper, we present our results for NGC~604. Our analysis of the STIS data for 30 Dor is still in progress and will be presented elsewhere. We have successfully extracted spectra for 49 stars and have assigned spectral types for 40 of these stars. In Section 3, we present individual STIS/G140L stellar spectra and spectral classifications based upon the UV spectral classification scheme of Smith Neubig \& Bruhweiler (1997, 1999). We also compare our findings with previous photometric estimates of the number of W-R/Of stars in NGC~604. In Section 4, for all 49 stars, with and without spectral classifications, we give UV fluxes and measured photometric magnitudes obtained from archival HST Wide Field Planetary Camera (WFPC and WFPC2) imagery. We combine the derived spectral types and photometry to determine the interstellar extinction law unique to NGC~604 in Section 5.  We place each star on the log(L$_{*}$/L$_{\sun}$)-Log(T$_{\rm eff}$) plane in Section 6. We further superpose the theoretical evolutionary tracks appropriate to the NGC~604 stars upon this H-R diagram. We use the combined observational and theoretical data to estimate evolutionary ages and the IMF for the massive stars in NGC~604. We explore the implications of this study on the spectral modeling of the total UV light of starburst spectra in Section 7. Finally, in Section 8, we present our conclusions and summary.

\section{THE OBSERVATIONAL DATA}

Our observational goal is to combine STIS/G140L spectra and HST photometry and obtain positions of the O \& B stars in the upper H-R diagram of NGC~604. Because of the well-known degeneracy of photometry both in the visual and in the ultraviolet in determining stellar effective temperatures for hot O stars, one must use stellar spectroscopy to obtain reasonably accurate effective temperatures and bolometric corrections. We use the broad- and medium- band HST/WFPC and WFPC2 filters to derive magnitudes and photometric colors, which are then used to determine the actual interstellar extinction law and extinction for the stars.  

\subsection{The STIS/G140L Spectral Data}

We acquired a single STIS two-dimensional spectral image using the G140L grating and the FUV-MAMA detector of the brightest concentration of early type stars in NGC~604. The 2$\arcsec$ wide aperture sampled roughly a 25$\arcsec$ x 2$\arcsec$ region in NGC~604.  The resulting UV spectral image provided a spectral resolution of approximately 2\AA\ and wavelength coverage of 1150-1730\AA. The wavelength coverage varies slightly depending upon the relative positions of the observed stars in the 2$\arcsec$ aperture. In Figure 1, we present the outline of the aperture with respect to the stars sampled in NGC~604. This is not an image of the aperture at the time of the observation, but our reconstruction based upon modeling of HST/WFPC2 imagery acquired with the F170W filter. The actual STIS spectral image is presented in Figure 2. An additional discussion of that image is presented in Paper 1.

We had designed our observations to have an aperture position and orientation to obtain the maximum number of O \& B stars within the aperture.  However, upon inspection of the data, we discovered that the actual aperture position (RA = 1$^h$~34$^m$~32.51$^s$ and $\delta$ = 30$^o$~47$\arcmin$~07.9$\arcsec$) was displaced by $\Delta$RA = -0.03$^{s}$ and $\Delta\delta$ = +2.6 $\arcsec$ from our planned position.  Figure 1 denotes the actual aperture field-of-view (FOV).

Our goal was to obtain a high pointing accuracy.  In an attempt to accomplish this, we performed an offset star acquisition.  We chose an offset star that appeared both on the Guide Star Survey plates and on a WFPC2 image containing the target FOV.  We painstakingly determined our slit location and orientation on this WFPC2 image.  According to the HST Instrument Handbook, a pointing accuracy of 0.01 $\arcsec$ is expected following a target acquisition exposure. 
Obviously, our incurred error was much greater than predicted.  The reason for this pointing error is unexplained.  Nonetheless, we did obtain a large number of O and B stars approaching that of our original expectation. We further compared the total STIS flux level summed over the length of the aperture at 1300\AA\ with that from International Ultraviolet Explorer (IUE) exposures of NGC~604 obtained with a 10$\arcsec$ x20$\arcsec$ aperture. The integrated STIS 1300\AA\ fluxes are roughly 60\% of the IUE fluxes and show that our STIS exposure samples a major fraction of the O \& B stars in NGC~604.
 
\subsection{HST Photometric Data}

We have incorporated much of the relevant archival HST broad- and medium-band imagery in our analysis. We include our own photometric results based upon HST/WFPC2 imagery obtained with the F170W, F336W, F547M, and F814W filters. We also adopt the pre-COSTAR WFPC F439W photometric results obtained by Drissen et al. (1993). However, since these results are based upon pre-COSTAR abberated PSF imagery, some of the deduced B-magnitudes should be viewed with caution. In particular, the stars 530D and 938 (1 of 2) have very unusual B-V colors (See Table 2). Hunter et al. (1995) have given derived stellar magnitudes from both WFPC2/F547M and F814W imagery for individual stars in NGC~604. However, they concentrated on stars of intermediate mass and presented few magnitudes and colors for our observed stars. We have used the Hunter et al. magnitudes for those stars in common to our study and reanalyzed the F547M and the F814W images to determine magnitudes and colors for the remaining stars. Our derived magnitudes are in good agreement with those of Hunter et al. and represent an external check for the accuracy of our photometric magnitudes for stars not given in Hunter et al. (1995).

A listing of the combined STIS and WFPC2 datasets used in our analysis is given in Table 1. Column 1 records the HST instrument that obtained the dataset. Column 2 contains the filter or grating appropriate for the dataset. Column 3 lists the central wavelength of the filter or grating. And, Column 4 gives the actual dataset name. In the case of the WFPC2 F336W, F547M, and F814W data, the stellar images relevant for the photometry lie on the WF3 chip.  However, the F170W dataset (again, see Figure 1) has our observed stars on the higher spatially sampled PC1 chip (0.046$\arcsec$/pixel versus 0.01$\arcsec$/pixel for the WF chips). See the WFPC2 Instrument Manual (Biretta et al. 2000) for a complete discussion of the instrument.

\section{Data Reduction} 

We have described the procedure for data reduction of the STIS 
spectral imagery in Paper 1 (Miskey \& Bruhweiler 2003). We will not duplicate that presentation here, and the reader is referred to that paper.

We have obtained the HST/WFPC2 images from the STScI archive, which were 
processed using the on-the-fly calibration. For all the WFPC2 filter 
data, two exposures are available which facilitated the removal of 
cosmic-ray hits. The accompanying data quality images in the HST/WFPC2 datasets were then used to mask or remove other bad or questionable data to produce relatively "clean" images for photometric analysis. The two exposures for each field were summed to produce the final reduced WFPC2 imagery. 
We have used the codes in {\it HSTphot}, developed by Andrew Dolphin (cf. Dolphin \& Kennicutt 2002) to obtain photometric magnitudes. \footnotemark
\footnotetext{The codes comprising {\it HSTphot}, as well as instructions for running them, can be readily obtained through the World Wide Web at 
(http://www.noao.edu/staff/dolphin/hstphot/).} 

\subsection{The Extracted UV Spectra}

In Figure 3a through 3l, we present the flux and wavelength calibrated G140L UV spectra of eleven O stars and one B star extracted with our software. In Table 2, we give the measured stellar continuum fluxes at 1300 and 1700\AA\ for the 49 stars for which meaningful spectral extractions were possible. We also present the 40 derived spectral types for stars where the S/N is sufficient to do reliable classification. (The data files containing the extracted spectra are available and can be readily obtained from the authors.) Each stellar spectrum is identified by its y -position in pixel units in the STIS spectral image. For groups of stars with overlapping spectra, their identifications are supplemented with alphabetical designations (i.e. 578A, 578B, 578C). The y-positions are marked along the vertical axis of Figure 2. Thus, one can easily associate the extracted spectrum with the unextracted spectrum on the STIS two-dimensional image. The extracted spectra are all of high quality, where the stars range in V from approximately 17.1 to 22.2 magnitudes. In the extraction process, as discussed in Paper 1, the strong interstellar absorption lines serve as fiducials for the wavelength and sensitivity calibrations for the stars. A more refined calibration is necessary since the observed stars are positioned randomly throughout the 2" wide aperture as depicted in Fig. 1. Since the interstellar lines are sharp, they are easily discerned in all the spectra. This enables us to determine where the star is positioned in the aperture and derive a proper wavelength correction. In all cases, the derived shifts agree with the implied spatial offsets for the stars denoted in Fig. 1. The interstellar lines are marked for the stellar spectrum shown in Fig. 3a. These lines include the N I triplet $\lambda$$\lambda$1198,1199,1200, Si II $\lambda$1260, the interstellar blend near 1300\AA\ due to O I $\lambda$1302 and Si II $\lambda$1306, plus C~II $\lambda$$\lambda$1334,1335, and Si II $\lambda$1526. In spectra where high signal-to-noise compensates for the rapid falloff in sensitivity at longer wavelength, Fe II $\lambda$1608 and Al II $\lambda$1670 are also seen. A full discussion of the stars with derived spectral types follows in Section 4.

\subsection {Derived Stellar Magnitudes \& Fluxes}

In Table 2, we present photometric data for the 49 stars with measured fluxes at 1300\AA\ and 1700\AA.  For all stars, we give derived magnitudes where possible using the WFPC2/F336W, F547M, and F814W imagery. In addition, the table includes the WFPC F439W B-magnitudes from Drissen et al. (1993). At the bottom of Table 2, one finds pertinent comments.

The WFPC2/F547M, WFPC2/F814W, and WFPC/F439W magnitudes are important for deriving E(B-V) values for the stars. In deriving these magnitudes, one must consider the contamination by nebular emission lines from the surrounding H II region and stellar line emission from W-R stars. This emission can lead to systematically larger flux estimates and erroneous E(B-V) values. The strong nebular contaminants are the [O III] $\lambda\lambda$4959,5007, and the Balmer lines H$\alpha$ ($\lambda$6563) and H$\beta$ ($\lambda$4861), as well as the satellite [N II] $\lambda\lambda$6548,6583 lines to H$\alpha$. The corresponding strong stellar emission are the Balmer lines, He II $\lambda$4686, and the carbon-nitrogen emission complex at 4650\AA\ in luminous O and W-R stars. To minimize the effects of these lines, we have used the F547M filter with its smaller effective bandpass to estimate the stellar V-magnitude. This filter has a spectral bandpass sufficiently narrow that the strong nebular and stellar emission lines have negligible contribution (cf. Biretta et al. 2000; Bruhweiler et al. 2001). 

As pointed out in the study of Drissen et al., the F439W filter bandpass of the pre-COSTAR WFPC ($\lambda$$_{\rm c}$ = 4350\AA; $\Delta\lambda$ = 510\AA) can be contaminated by nebular plus stellar H$\beta$ and He II 4686\AA\ emission. An additional complication arises since the WFPC, unlike the WFPC2, did not have corrective optics to compensate for the spherical aberration of the HST primary mirror. Consequently, the WFPC/F439W data do not have the spatial resolution of imagery obtained with the WFPC2. Thus, in some cases, individual stars may not be resolved in the F439W data. 

In determining the reddening of stars in NGC~604 at wavelengths longward of the Balmer jump, we give the reddening deduced from the WFPC2/F547M and F814W data higher weight than reddening estimates that incorporate the WFPC/F439W filter data, for which the deduced B-magnitudes may have been affected by the abberated PSF of the WFPC. The strongest nebular emission line sampled by the F814W is likely [Ar~III]~$\lambda$7136. Yet, the [Ar~III] flux is probably only 5\% of that of H$\alpha$ if comparisons with 30 Doradus are valid (Dufour, Shields, and Talbot 1982). This should be the case since 30 Dor appears to have elemental abundances similar to those in NGC~604. We thus expect the colors and reddening deduced from the WFPC2/F547M and the F814W data to have negligible contamination from nebular emission.

Special consideration is given to the WFPC2/F170W data. Of the 
WFPC2 data used in our analysis, only the F170W stellar imagery comes from the PC chip of the WFPC2. All other photometric data for the stars observed by STIS were derived from the WF3 chip. Because the angular pixel size of the PC chip is 0.046 $\arcsec$ or half that of the WF chips, the stars are better sampled leading to better positions and potentially better effective spatial resolution. Moreover, the central wavelength of the F170W filter ($\lambda_{\rm c}$ $\sim$ 1749\AA) corresponds to the higher wavelengths in the spectral range covered by the STIS spectra. Thus, the F170W derived fluxes should have an extremely good correspondence with the fluxes longward of 1500\AA\ for the STIS/G140L stellar spectra. 

Originally we had planned to use the F170W photometry as a means to test the reliability of our spectral extraction process (Miskey \& Bruhweiler - Paper 
1). We thought that the F170W imagery should be particularly useful for determining the accuracy of flux extractions of stars that have small spatial separations. But, the derived F170W fluxes proved not very accurate. The CCD windows of the WFPC2 slowly accumulate contaminants and produce a strong degradation of light throughput most notable in the UV (Holtzman et al. 1995). To minimize degradation, the CCDs are warmed (annealed) periodically at roughly monthly intervals. Despite the annealing, Holtzman et al. report variations across the F170W chip of 10-20\% after 30 days indicative of non-uniform deposition of contaminants.  At the same time, a fit to ascertain the degradation slope indicates a 0.2 magnitude decrease in F170W fluxes over the same $\approx$ 30 day period.

Nonetheless, we still underwent our own evaluation using the STIS spectra and the archival F170W imagery.  We have compared the deduced fluxes at 1730\AA\ for the STIS spectra with those inferred from the F170W photometry using the Holtzman et al. (1995) flux conversions for the PC1 chip. In theory, this comparison should yield a rough gauge of the consistency of the STIS and F170W derived fluxes. In some cases, we have extrapolated the 1730\AA\ fluxes from the STIS spectra. In all cases, the fluxes from the STIS data are higher than fluxes from the WFPC2/F170W data. Based upon a comparison of 10 stars where we have reliable STIS flux estimates, we find a relation between the STIS and F170W fluxes, F$_{\rm STIS}$ = 1.3825(F$_{\rm F170W}$) + 3.73 x10$^{-16}$ 
ergs~cm$^{-2}~$\AA$^{-1}$. There is significant scatter about this relation, which presumably reflects the variable rate of contaminant deposition onto the PC1 chip.  Because of the fall off in sensitivity for the G140L configuration at the long wavelength end of its range, the STIS fluxes in the 1700\AA\ region are less accurate than fluxes at shorter wavelengths.  Yet, the trend and basic results are valid. The small flux offset of 3.73 x10$^{-16}$ ergs~cm$^{-2}$~\AA$^{-1}$ seems to be real, but we have no explanation for its origin. Due to these problems, we do not give F170W fluxes or magnitudes here.

We have used the ratios between the STIS/G140L and the WFPC2/F170W fluxes to assess how well our spectral extraction algorithm disentangles the overlapping spectra of stars that are very close together. For example, the STIS and corrected F170W flux ratios for stars 578A, B, and C are roughly the same, namely 3.2: 2.6: 1. This is further evidence that our extraction algorithm discussed in Paper 1 is properly extracting spectra and yielding reliable fluxes for stars that are marginally spatially resolved in the STIS/G140L dataset.

\section{The UV Stellar Classifications.}

As mentioned previously, the UV spectral types derived for the NGC~604 stars are based upon the UV classification scheme of Smith Neubig \& Bruhweiler (1997; 
1999). This classification scheme was built upon analyses of IUE spectra of O and B stars in the Galaxy, LMC, and SMC. The 12 representative UV spectra presented in Figure 3 clearly typify an extremely young population of O stars. Three of the stellar spectra shown are O4 stars and must have ages less than 3 Myr.  These stars are 867A, 867B, and 117. Three additional stars are of spectral type O5, namely stars 825, 775B, and 690A. All the O4 stars exhibit definite O V $\lambda$1371 absorption, which is a clear indicator of either O4 or extremely rare O3 stars. The identified O4 stars also have the strongest N~V$~\lambda$$\lambda$1238,1242 P Cygni features of any of the stars studied here, including exceedingly strong emission lobes. Only in 690A and 578B, with spectral types of O5 III and O9 Ia, are the N~V comparable to these O4 stars. In our previous IUE results (Smith Neubig \& Bruhweiler 1997; 1999), we found that N~V remains strong in O supergiants, while near the main sequence the N~V decreases in strength until it is not detectable in stars later than O6. This behavior should hold true for the STIS spectra, even though the STIS/G140L resolution is 3x better than that of IUE. In short, the N~V is strongest at high temperatures and demonstrates a luminosity dependence.

The wind features of Si IV $\lambda$$\lambda$1393,1402 are well-established luminosity indicators. The Si IV wind profiles are quite strong in O9 
supergiants and decrease in strength toward the earlier O and later B spectral types. In main sequence O stars at G140L resolution, stellar Si IV should be absent. The classifications in Figure 3 are largely consistent with this expectation. The only possible exception is the O4 Ia classification for 867B, where the strength of the S IV, if one compensates for interstellar contribution, is weak or absent. The spectral classification also reflects the relative strengths of the N V and C IV, which are the primary UV luminosity indicators in the earliest O stars (Smith Neubig \& Bruhweiler 1999). However, the implied intrinsically weak Si IV might suggest that 867B is an unrecognized binary system.

\subsection{The Presence of Wolf-Rayet Stars}

For stars with UV classifications, we find four stars in the 
STIS data that have been identified as Of or W-R stars in Drissen et al. (1993). Of these, only three reveal UV line emission in the STIS data other than that due to the wind profiles of N V, C~IV, and Si IV. This additional emission indicates Of/W-R stars. These stars are 867B, 867A, and 578A. All three stars have very pronounced He II $\lambda$1640 emission. The stars 867B and 867A, respectively, are the UV brightest stars sampled in our STIS exposure of NGC~604. Although stellar He II emission is considered a signature of Wolf-Rayet stars, because of the apparent high luminosity of 
867A and 867B, these stars may be Of stars rather than W-R stars. Both the 
He II emission in 867A and B are accompanied by C~IV P Cygni profiles with strong emission lobes, which may also imply that these stars are WC-type W-R stars.  In 578A, no C~IV emission lobe is detectable, but there is possibly weak N III] 1485\AA\ emission (See Fig. 3j). We further note that the visual spectrum of the stellar aggregate containing 578A also shows He~II $\lambda$4686 (Drissen et al. 1990) with no hint of carbon emission. In addition, the presence of any He~II in an O9~II star is highly unusual. Thus, the evidence strongly suggests that 578A is a true WN star. In Figure 3, if we ignore the He~II, we give both an O9~II classification based upon the UV classification criteria alone, plus a WN classification (in parentheses) based upon the optical data. These emission line characteristics clearly indicate that 578A is a WN-type W-R star.

We have also compared our results with the photometric study of Drissen et al. (1993), who used HST photometric data to estimate the number of W-R stars in NGC~604 and NGC~595. They used the ratio of the F439W and the F469N filter fluxes from the WFPC to detect flux excess due to He~II  $\lambda$4686 stellar emission. Notable He II $\lambda$4686 emission, as is He II $\lambda$1640 emission in the UV, is the signature of a W-R or Of star. Many of the stars sampled by Drissen et al. are common to those observed in our STIS exposure. Drissen et al. identified 14 W-R/Of candidate objects in NGC~604 using their photometric method.

All three of the O stars with He II $\lambda$1640 emission in the STIS data were also identified as W-R/Of star candidates by Drissen et al. (1993). The STIS stars 867B and 867A correspond to the Drissen et al. stars \#81 and \#84, respectively. These stars are the fourth and fifth brightest stars in 
NGC~604 in terms of B-magnitude, with B = 17.31 and 17.65 as given in Table 2 of Drissen et al. These stars are also denoted as WR2b(867B) and WR2a (867A) in that table. The stars, 578A, 578B, and 578C, in our study correspond to stars \#202, \#200, and \#196, in Drissen et al. Although Drissen et al. list both \#202 (WR4a) and \#196 (WR4b) as W-R/Of candidates, we find definite He~II UV emission only in 578A (\#202). 
One might argue that the high continuum level at 1640\AA\ in 578C (Fig. 3h) is due to He~II. However, it appears that this is largely due to a window of low line opacity that is bracketed by strong Fe~IV absorption at shortward wavelengths and by strong interstellar Al II $\lambda$1670 at longer wavelengths. Strong 1600-1630\AA\ Fe~IV absorption is also apparent in 578B (Sp. Type = O9 Ia) and in 690B (B0 Ib), both of which are similar in spectral type to 578A. Moreover, there is no hint of any other strong UV emission, especially from nitrogen or carbon ions, in 578C. Despite this, we cannot completely rule out the possibility of weak He~II $\lambda$1640 in 578C. 

Finally, 690B in the STIS data is also listed as \#161 or WR12 in Drissen et al. and a W-R/Of candidate. An examination of Fig. 3f clearly shows no trace of He II $\lambda$1640 emission. Thus, we find no evidence that 690B is either a W-R or Of star. Out of the five W-R/Of star candidates common to the two studies, we find three objects with detectable He~II $\lambda$1640 emission. None of the other stars show any hint of UV emission in He~II.

\section{Interstellar Extinction.}

We have attempted to derive a preliminary interstellar extinction law for 
NGC~604 and then apply that law to the complete observed flux distributions to derive estimates to the interstellar extinction in the V and B-bands. These are later used (See 5.3) to derive the absolute luminosities for the observed stars in NGC~604.

A lower limit to the extinction is imposed by the inferred Galactic extinction. The Galactic extinction, deduced from a reprocessed composite of the 
COBE/DIRBE and IRAS/ISSA maps in Schlegel et al. (1998) as given in the NASA 
Extragalactic Database (NED), is A$_{V}$ = 0.151. This value is slightly higher than that deduced from the H~I maps of Burnstein \& Heiles (1982), which assume a constant gas-to-dust ratio. We have corrected for Galactic extinction by applying this A$_{V}$ and the Savage \& Mathis (1979) extinction curve to both the UV STIS spectra and the deduced fluxes from the HST imagery acquired with the F336W, F439W, F547M and F814W filters. 

We then derive an extinction curve that is applicable to NGC~604. To accomplish this, we adopt the Kurucz (1979) model atmosphere flux distributions appropriate to the deduced spectral types and luminosity classes for the stars. For $\lambda$ $\geq$ 3000\AA, we assume the Savage \& Mathis extinction law along with the normalization, A$_{V}$ = 3.1(E(B-V)). This relation is used to derive an A$_V$ intrinsic to NGC~604 such that the dereddened spectral shape of the dataset, previously corrected for Galactic extinction, reproduces the shape of the Kurucz flux distribution at visual wavelengths. Then, the required attenuation, A$_{\lambda}$, to match the intrinsic slope of the Kurucz flux distribution, is determined for the UV over the range $\approx$ 1200-1650\AA. This is done repeatedly for a number of stars, which yields the mean UV extinction curve for NGC~604 (See Fig. 4). Even though the UV data typically span a larger wavelength range (1150-1730\AA), UV data outside the 1200-1650\AA\ range were assigned low weight due to the falloff in MAMA/G140L sensitivity at these extreme wavelengths.  We have no data covering the well-known 2175\AA\ interstellar extinction bump, and therefore can say nothing about its presence in the NGC~604 extinction curve. 

The resulting extinction curve in Figure 4 is much flatter than the average Galactic extinction curve of Savage \& Mathis (1979). The curve that we have derived is only an approximate curve since the number of stars with reliable fluxes over a wide enough wavelength range in the visual is small. The derived E(B-V) is subject to the uncertainties in the deduced absolute fluxes from the respective filters. These flux uncertainties could be as large as 10\% in some cases. To a lesser degree, the extinction curve is sensitive to the choice of model atmosphere flux distribution used. This is further compounded in that the Kurucz models do not cover the full range of T$_{\rm eff}$ and log(g) for the stars. This is especially the case for the hottest and most luminous O stars. Also, the Kurucz model atmospheres are calculated for solar metallicity, assuming plane-parallel geometry and LTE. However, the shape of the flux distribution versus temperature and gravity does not appear to vary significantly for the hottest O stars.  None of these uncertainties appear to alter the basic conclusion, namely the shape of the NGC~604 extinction curve is flatter than the Savage \& Mathis (1979) curve at shorter wavelengths.

In choosing the proper model atmosphere flux distribution, we originally adopted the spectral type - effective temperature scale and bolometric corrections for O stars in the LMC of Fitzpatrick \& Garmany (1990). As we have previously stated, the LMC has a metallicity comparable to NGC~604. In the hotter O stars and those evolved away from the main sequence, the grid of Kurucz models do not extend to low enough gravities. Thus, we have used the Kurucz model of lowest possible gravity and effective temperature that most closely matches the effective temperature implied by the UV spectral classification. We also calculate normalization factors that would exactly overlay the theoretical Kurucz spectrum on the dereddened observed flux distribution. These normalization factors, if the unit conversions are made for flux between observations and model atmospheres, and if the dereddening is correct, are actually the geometric dilutions, D= $\pi$(R$_{*}$/d)$^2$, where R$_{*}$ is the stellar radius and d is the distance to NGC~604. 

We use the derived NGC~604 extinction curve to deredden all the stars with UV spectral types and determine the effective E(B-V). These stars all have reliable UV fluxes, spectral types, and at least one stellar flux measurement from either WFPC or WFPC2 imagery at wavelengths longward of 3000\AA. For nine stars without accurate WFPC or WFPC2 photometry, we derive the reddening by applying the extinction correction to the STIS spectra to match the Kurucz flux distributions. We use these reddening values to deduce an extinction correction in the V-band. We do this by dereddening the observed flux distribution such that the dereddened spectral slope and shape mirror that of the Kurucz model atmosphere closest to the implied effective temperature and gravity for the assigned spectral type.  The implied E(B-V) for each star is given in Table 3.  The stars with UV spectra and optical photometry have E(B-V) uncertainties of 0.05 magnitudes, while the stars without optical magnitudes in Table 2 have uncertainties of $\sim$ 0.1 magnitudes.

We have further compared the Kurucz models with those of the WMBasic O star grid that are available by L. J. Smith\footnotemark
\footnotetext{The grid of WMBasic model atmospheres can be obtained through the World Wide Web at (http://www.star.ucl.ac.uk/~ljs/starburst.html).}
(Also see Smith et al. (2002) and Pauldrach et al. (2001)). Specifically, we compared the Kurucz models with the WMBasic models calculated for z=0.008, comparable to the abundance of NGC~604. In general, the comparisons suggested that our deduced extinction was not affected significantly by the choice of models. As we mention above, Kurucz models for O stars are generally limited to higher gravity (log(g) =4.0), more appropriate for main-sequence and relatively unevolved stars. We found that the Kurucz main-sequence model flux distributions used, when normalized at 5500\AA, agreed quite well longward of the Balmer jump with the WMbasic models, but yielded UV fluxes slightly lower than the WMBasic models for the same gravity. The Kurucz log(g) =4.0 models are in very good accord with the WMBasic models for log(g) =3.5 at 35,000K. Our detailed comparisons imply that the extinction derived using Kurucz models will have slight systematic differences with the WMBasic models. Our approach using the Kurucz models would provide extinctions that are slightly smaller for main-sequence O stars, while slightly larger for supergiants than would the WMBasic models.

Another problem is that each grid of model atmospheres requires its own 
T$_{\rm eff}$-calibration with the observed O star spectral energy distributions.  To better address this issue, we are in the process of using the O star model atmospheres calculated using the CMFGEN code (Hillier \& Miller 1998)to reproduce the UV through visual flux distributions and extinction law of O stars in 30 Dor. We expect that the results will have direct applicability to the stars studied here.

\subsection{Implied Agreement between Geometric Dilutions and Spectral Types}

We use the calculated geometric dilutions as a consistency check for our deduced extinction values, assigned spectral types, and luminosity classes. Specifically, we compare the calculated factors, D, with the UV spectral types (effective temperatures) and luminosity classes in Figure 5. Our derived luminosity classes are in good agreement with the derived geometric dilutions. This gives some indication of the accuracy of the assigned luminosity classes. The most problematic stars are also the ones that have the most uncertain spectral types as listed in Table 2 (also see Table 4). We also, in Figure 5, show lines of constant stellar radius delineating stellar radii, R$_{*}$ = 10 R$_{\sun}$ and 25 R$_{\sun}$. Since the distance, d, is a constant, the value of D is proportional to (R$_{*}$)$^2$ and the stellar luminosity for each spectral type. These radii provide a comparison for stars of different effective temperature and luminosity class. 

We emphasize that the spectral types are derived independently of the geometric dilutions. We do not use the calculated factors, D, to refine our spectral typing. It is very important that these are derived separately, since any discrepancies might signal the presence of unresolved stellar clustering. Because of the good agreement between the derived luminosity classes and the derived geometric dilutions, we do not think that unresolved aggregates are important. However, we return to this problem later when we discuss the most luminous stars.  

\section{Stellar Parameters and the log(L$_{*}$) - log(T$_{\rm eff}$) Plane}

One of the most important aspects in establishing a reliable distance scale within the Local Group is determining the distance modulus for the LMC. We have adopted a distance modulus of 18.5 for the LMC, which reflects the mean of values determined from eclipsing binaries (Fitzpatrick et al. 2002; Guinan et al. 1998; Guinan 2002) and Cepheids (cf. Mould et al. 2000). The latest results, published and unpublished, for the eclipsing binaries suggest 18.4 to 18.5 (Guinan 2002), while the Cepheids imply 18.5 or possibly higher (Madore \& Freedman 1991; Mould et al. 2000). Applying these corrections, we have modified slightly the distance modulus of Freedman et al. (1993) and adopt a distance modulus of 24.64 $\pm$ 0.09 magnitudes for NGC~604. We use this in deriving the stellar luminosities for our observed stars. 

The adopted stellar effective temperatures, luminosities, and implied E(B-V) are presented in Table 3 for stars with determined spectral types. Also in Table 3, we present corrected effective temperatures for stars on the main sequence based upon the revised O star calibration presented in Martins, Schaerer, \& Hillier (2002). Since this corrected scale is only available for main sequence stars, we give adjusted effective temperatures only for stars classified as luminosity class V. We did not adjust the luminosities since Martins et al. conclude that the luminosities are $\leq$ 0.1 dex lower. 

In Figure 6, we plot parameters for all stars in Table 3 on a luminosity-effective temperature diagram.  The evolutionary tracks from Meynet et al. (1994), representative of stars with metallicity Z = 0.004, are superposed. In Fig 6a, we have plotted the effective temperatures and luminosities without the corrections suggested by Martins et al., while, in Fig 6b, we have plotted the corrected values relative to the evolutionary tracks of Meynet et al. The uncertainty in the stellar effective temperature is largely due to the quantized value of the effective temperature given to each spectral classification as presented in Fitzpatrick \& Garmany (1990). The error-bars in luminosity reflect the uncertainties in the distance of NGC~604 and bolometric corrections linked to uncertainties in stellar effective temperatures.

Inspection of Figure 6a reveals a clustering of O stars at lower luminosity that is below the ZAMS. The uncertainty in the distance modulus for NGC~604 is too small to explain away the problem of having stars below the theoretical ZAMS. Moreover, our implied extinction corrections are uniformly low and derived with an extinction curve unique to NGC~604. Increasing the extinction, and hence the deduced stellar luminosities, could move these stars above the ZAMS, but would also increase the luminosities of such objects such as 867A, 867B, and 578C, which already appear to be exceedingly bright, with inferred masses M $\geq$ 120 M$_{\sun}$. Given our derived extinction law for NGC~604, we think that significantly higher extinctions for the stars of NGC~604 are very unlikely. 

If we adopt the revision to the main sequence O star effective temperature calibration of Martins et al. (2002), then the problem of having stars below the theoretical ZAMS is largely removed, as is seen in Figure 6b. We now find that only one star, 745B, necessarily falls below (or to the left of) the ZAMS defined by the Meynet et al. (1994) evolutionary tracks. The assigned spectral type of this star is highly uncertain, as indicated in Table 2. It is one of the faintest stars with a spectral type given in Table 2 and plotted in Figure 6, and has a high degree of spectral overlap with 745A (See Figures 1 \& 2). Thus, we suspect that the spectral type for 745B may be in error.  The agreement of the positions of the unevolved O stars and that of the ZAMS give strong support to the validity of the revised O star effective temperature scale of Martins et al. (2002).

A comparison of the evolutionary tracks and deduced effective temperatures and luminosities of the intrinsically brightest O stars implies that all of these stars are slightly evolved and away from the ZAMS with ages $\sim$~3~Myr. This age may indicate the most recent burst of star-formation. However, the paucity of evolved O stars does not provide meaningful statistics, and even more recent, or ongoing, star-formation is highly likely.  In what follows, we assume that massive star formation has only been occurring within the last 3~Myr. This is particularly important in our discussion of the NGC~604 IMF.

Of the 40 STIS/G140L UV sources with spectral types, only three are found to be B supergiants; the remaining are O stars (see Table 2). The latest main sequence spectral type is O9. Given that a star spends roughly 90\% of its lifetime near the main sequence, this (B star)/(O star) ratio is not inconsistent with expectations for either a short episode (possibly a burst) of star formation that was initiated within the last 3 Myr, or for an ongoing, constant star formation rate. Thus, the statistical significance of three B supergiants should not be overstated for a region like NGC~604, which has undergone recent star formation.

One needs to be extremely careful when comparing evolutionary tracks with deduced effective temperatures and luminosities of the hottest and most luminous O stars. Since the hottest O stars have a significant fraction of their emergent flux arising in the inaccessible extreme-UV shortward of the Lyman-jump, the effective temperature scale and bolometric corrections for these stars rely heavily on a calibration using model atmosphere calculations.  It has long been known that the O stars, especially the hottest objects, have pronounced non-LTE effects and enhanced mass loss. In addition, spherical instead of plane-parallel geometry becomes more important. Even stellar rotation plays a basic role in determining the flux distribution of these stars. To properly account for these effects requires much more sophisticated stellar atmosphere modeling that abandons hydrostatic equilibrium in favor of a more realistic hydrodynamic flow representation.  

Indeed, modeling efforts have gone far in reproducing the observed characteristics of O stars in and outside of our Galaxy. Currently, the effective temperature scale and bolometric corrections for the earliest O stars are somewhat uncertain. They are dominated by a variety of systematic errors that stem from problems in obtaining an accurate description of the physics and opacity sources that affect the spectral characteristics of O stars. A discussion of the problems can be found elsewhere (Vacca et al. 1996; Martins, Schaerer, \& Hillier 2002). For example, the recent results by Martins et al. (2002), which include non-LTE line-blanketing and the effects of the stellar wind, lead to effective temperatures approximately 4000K lower in the earliest O stars, with changes in bolometric corrections leading to systematically lower luminosities. The effects of line-blanketing still are important for stars in the SMC, which have metallicities lower than those of NGC~604. Thus, the reader must be aware that these additional systematic uncertainties are not included in the error-bars in Figure 6.  The sizes of the systematic uncertainties are largest for the O4 stars and much smaller for the cooler stars. 

The stellar evolutionary tracks of Meynet et al.(1994), or any others that might be used,  may also contain systematic errors, especially in effective temperature. These tracks are based upon stellar evolutionary codes that attempt to give the best representation for interior opacities and nuclear reactions. They are not designed to yield accurate modeling of the opacities in the stellar photosphere and winds where the emergent flux distribution is formed. To first order, conservation of flux implies that the total luminosity is equal to the energy generation rate in the interior. Since these stars only increase their luminosities slightly as they evolve away from the ZAMS in the log(L$_{*}$) - log(T$_{\rm eff}$) plane, the total luminosity should provide a reasonable indicator of the initial main sequence mass. It is unclear if systematic errors exist between the deduced effective temperature scale based upon model atmospheres and that derived from the evolutionary codes. (For a more detailed discussion on determining the effective temperature scale for the evolutionary tracks used here, see Schaller et al. (1992).) Ideally, one would like to have a completely consistent physical treatment to describe both the O star evolution and model atmospheres. This is currently not available. 

Despite all of this, it is very encouraging that one gets such good agreement between the theoretical ZAMS and the model atmosphere derived effective temperatures and luminosities of the later, cooler O main sequence stars in Figure 6b. The uncertainties in the distance modulo for nearby galaxies appear to be small enough to actually test the agreement of the newly revised O star temperature scale and the predicted effective temperatures for the ZAMS from the evolutionary models.

\section{The IMF for the O stars in NGC~604.}

The number of O stars in each mass range can be derived from a straightforward examination of the distribution of O stars within the mass interval defined by the Meynet et al. (1994) evolutionary tracks for stars with metallicity, Z =0.004 where Z$_{\sun}$ =0.02.  We then compare the observed number of stars in each mass range with the expected number of stars assuming a constant $\alpha$ power law representation for the total number of stars, $N_{*}$, between limiting stellar masses, m$_{L}$ and M$_{U}$ using 

\[N_{*} = A\int_{m_L}^{m_U}m^{\alpha}dm. \]

In doing this, we first considered all the stars in Table 3. In ascertaining the slope of the IMF, we only consider stars with implied main sequence masses $\geq$ 20 M$_{\sun}$, because we think that we have a complete sampling of these stars in the 25$\arcsec$ x 2$\arcsec$ FOV sampled by the STIS aperture.  This provided a culled sample of 30 stars.  The best fit was for a uniform power law slope for the entire mass range of $\alpha$ = -2.3, which is the same as the Saltpeter (1959) IMF.  For example, the $\chi^2$ is 4.52 for $\alpha$ = -2.3 compared to 8.75 and 8.05, for $\alpha$ = -2.0 and -2.7, respectively. The case for $\alpha$ = -2.7 corresponds to the power law slope for the most massive stars in the Miller \& Scalo (1979) IMF for the local solar neighborhood.

Perhaps the most important aspect of the derived stellar masses for the luminous O stars is that two, possibly three objects, 867A, 867B, and 578C, have implied masses above 120 M$_{\sun}$. Previous studies of O stars in the Galaxy and the Magellanic Clouds have deduced stars with M$_{*}$ $\sim$ 60- 85M$_{\sun}$ (Massey et al. 1995). Massey et al. have found the inferred upper mass limit of most stellar clusters is more statistical and not a true physical upper limit. The most massive stars tend to be found in the larger associations. In addition, further work by Massey and his colleagues (Massey et al. 1996) found M33 rich in massive stars and several intriguing objects in NGC~604 that warranted further study. Thus, instead of dismissing the luminous objects in NGC~604 as compact clusterings of O stars, we have attempted to investigate in more detail whether these objects are single or possibly compact clusters of stars.  We have closely examined both the STIS spectral imagery and the more highly spatially sampled PC imagery obtained with the F170W filter and the WFPC2.  The STIS spectral extractions and WFPC2 data clearly show no hint of multiplicity.

Our tests using 578A, 578B, and 578C clearly demonstrate that we can resolve spectra as close as 0.055$\arcsec$ perpendicular to the dispersion.  This angular distance corresponds to a separation on the plane of the sky of 0.21 pc at the distance of NGC~604. In Figure 7, we display the flux contours of the F170W PC imagery around the objects 867A \& 867B and also 578A, 578B, and 578C.  Again, close examination of the flux contours reveals no evidence of additional multiplicity in the 578 and 867 groupings. There is only a slight hint for additional flux for the two groupings of stars in Figure 7. If real, this low level flux would indicate stars of much lower mass and would not alter the conclusion that the stars in question are massive stars $\geq$ 120 M$_{\sun}$. The spectral characteristics then might suggest tight aggregates of W-R stars with concentrations smaller than the FWHM of the WFPC2/PC-chip point spread function. Certainly, some of the most luminous UV objects are still either unresolved tight clusters or massive binaries.

It is intriguing that these two tight groupings totaling five objects represent the five most luminous objects identified in our STIS spectral image.

These luminous objects have important implications for the IMF in NGC~604. We certainly cannot rule out the possibility that 867A and 867B, and possibly the objects comprising 578A, 578B, and 578C as well, are spatially unresolved close binaries. The unusual weakness of the Si IV in 867B, already noted in Section 4, might be evidence for binarity. If one maintains that these stars are, indeed, close binaries, then the UV spectral characteristics still argue for stars of high mass. This would still give stars with masses near 65M$_{\sun}$.  In deriving the slope to the IMF, we assume that we are seeing all the O stars, and that they were born in a recent episode of star formation.  As a result, we have not made evolutionary lifetime corrections that would be applicable under the assumption of continuous star formation. These lifetime corrections would be most important for the most massive stars. If we did assume that continuous star formation were present here, we would be pushed to shallower slopes for the IMF.

Spectroscopic observations at higher spectral resolution and infrared photometry have the potential to ascertain the nature of these luminous objects. Spectroscopic studies at higher spectral resolution might reveal that these are close binary stars of lower mass. Conversely, if these stars are stars of unusually high mass, they might exhibit unusual spectral and photometric variability or large infrared fluxes. This IR-excess would likely arise due to stellar instabilities occurring in a luminous blue variables (LBVs) or a star near or exceeding the Eddington limit. 

\section{Constraints on Starburst Modeling.}

We have further used our results to assess the reliability of population spectral synthesis of UV spectra of starbursts at greater distances where spectra of individual stars cannot be obtained. 
The spectra depicted in Figure 8 clearly show how the contributions of individual luminous stars determine the outcome of any spectral fitting to starbursts (cf. Gabel \& Bruhweiler 2002).  In Figure 8, we display the UV spectrum of the brightest UV source detected in our STIS spectral imagery, 867A. Along with this spectrum, we plot the sum of the UV flux for the 10 brightest objects seen in the STIS data. In addition, we depict the integrated UV light versus wavelength for the entire height, along the y-direction, for the STIS 25$\arcsec$ x 2 $\arcsec$ FOV. Although we have not properly aligned the spectra of the individual stars in the total integrated UV spectrum, several conclusions are evident. 

First, if a comparison is made between the integrated aperture flux and that for the ten brightest stars, it appears that the actual flux variations in the spectrum due to spectral features, seen for the full aperture, are determined by the spectral features of the brightest 10 stars. This is seen most dramatically for the observed flux variations in the N V, Si~IV, and somewhat in the C~IV 
P~Cygni profiles. This is also seen in the He II $\lambda$1640, the weaker photospheric Fe~IV features between 1600 and 1640\AA, and the Fe~IV absorption at 1450\AA.  Furthermore, the total UV flux from the aperture has the same spectral slope as that of the 10 brightest UV sources.

It is worthwhile to compare the results of this paper with those of Gonzalez Delgado \& Perez (2000), who examined IUE low resolution and ground-based spectra of NGC 604. Basically, their results, which also included population spectral synthesis of the integrated light of NGC 604 and constraints from photoionization modeling of the H II region emission lines, found that the stars were quite young with ages $\sim$ 3 Myr. Furthermore, the cluster could be fit by an instantaneous burst having an IMF of $\alpha$ = -2.35 or flatter and stars more massive that 80 M$_{\sun}$. These conclusions are almost identical to those presented here. Thus, the comparison of these two studies of NGC~604 provides strong confirmation of the technique of population spectral synthesis, especially when constraints on the ionizing flux from the nebular emission can also be used.

\section{Summary \& Conclusions.}

We have used our spectral extraction technique as developed in Paper 1 (Miskey \& Bruhweiler 2003) to extract individual UV spectra of 49 O and B stars in a single STIS/G140L exposure of NGC~604. We have derived 40 UV spectral types for these stars based upon our UV classification criteria (Smith Neubig \& Bruhweiler 1997; 1999) and combined the results with HST/WFPC and WFPC2 photometric colors to derive effective temperatures and luminosities of the observed stars. In the process, we have determined an extinction curve for NGC~604. The basic results of this work are as follows:

a.) The UV spectral types derived from the STIS/G140L spectra reveal a very young population of O stars. The presence of stars as early as O3 and O4, plus their positions on the L$_{*}$ - T$_{\rm eff}$ diagram relative to evolutionary tracks, indicate stars with characteristic ages of 3 Myr or possibly less.

b.) We identify three stars with prominent He~II $\lambda$1640 emission, characteristic of W-R/Of stars. One of these stars also exhibits weak N~III] $\lambda$1487 emission, which may signal the presence of at least one WN-type W-R star. 

c.) The deduced interstellar extinction curve for NGC~604 is flatter in the UV than that denoted by the Savage \& Mathis curve for our own Galaxy. We cannot say anything about the presence of a 2175\AA\ extinction bump, since we have no data spanning that feature.

d.) The derived extinction is low, slightly lower than that used previously by Drissen et al. (1993) in their photometric study of NGC~604.

e.) The placement of the O stars in the log(L$_{*}$) - log(T$_{\rm eff}$) diagram relative to the placement of the Meynet et al. (1994) evolutionary tracks indicate three objects with exceptionally high mass, above 120 M$_{\sun}$. Although the UV spectral characteristics imply high luminosity, we cannot definitively determine if these objects are single or unresolved multiple stars. 

f.) The comparison of the ZAMS with the lower luminosity main sequence O stars indicates that recent recalibration of the O star effective temperature scale by Martins et al. (2002) yields good agreement with evolutionary tracks in the case of NGC~604. This comparison is meaningful since the uncertainties in the distance modulo of nearby galaxies are now quite small thanks to recent work using the HST.

g.) The inferred slope of the IMF for 30 luminous O and B stars in NGC~604 is consistent with $\alpha$ $\approx$ -2.3. This result assumes that the observed stars were all formed in a recent episode of star formation.

h.) Comparisons with the integrated UV flux along the axis perpendicular to the dispersion indicate that the 10 brightest stars dominate the stellar UV spectral features produced by this starburst. This suggests that the integrated UV spectra of more distant starbursts can also be dominated by the spectra of a few stars and should be interpreted with extreme care.

We have taken every opportunity to provide consistency checks in our data analysis.  Our comparison of F547M magnitudes for those few objects in common with Hunter et al. (1994) show good agreement. We have used WFPC2/F170W data to check for systematic errors in flux levels for the extracted spectra of 578A, 578B, and 578C. These stars are barely spatially resolved and have substantial spectral overlap.  We have analyzed the derived geometric dilution factors to see if the spectral types, established extinction, and the inferred stellar radii are consistent. We have carefully examined the available imagery to ascertain, as much as possible, if any of our stellar objects were composed of multiple stars.

One important aspect of this work is investigating the formation of the most massive stars. Recent theoretical work (McKee \& Tan 2002a, 2002b) predicts that accretion with the presence of supersonic turbulence in molecular clouds can overcome the effects of radiation pressure to produce stars $\sim$ 100 M$_{\sun}$. However, there is no clear indication of what the theoretical upper mass limit for stars might be. The subject of very massive stars has always been quite controversial. Various investigators over the past several decades have postulated the existence of very massive stars (VMSs) well in excess of M$_{*}$ $\approx$ 120 M$_{\sun}$. These stars have been evoked as the Population III stars mainly to describe the rapid increase of heavy elements in the Early Universe (Heger \& Woosley 2001; Qian et al. 2002). The VMSs might also provide the rapid growth of massive black holes, which are presumed to be the engines that power quasars. Certainly, the presence of QSOs at z $\sim$ 5 demands rapid evolution and a strong constraint for forming AGNs.

If stars of inordinately high mass exist in the present day Universe, the best place to look for these objects are in the nearby starbursts where individual stars can be seen with instruments on HST or possibly through adaptive optics from large ground-based telescopes. We find in this study three objects that might have masses in excess of 120 M$_{\sun}$.  We emphasize that the search for super- or very-massive stars has been quite elusive, as the history of the compact core, R136a, of 30 Doradus dramatically shows (Savage et al. 1983; Weigelt \& Baier 1985; de Koter et al. 1997). 

This study of NGC~604 has provided intriguing, although tentative, results.  Recent results have greatly reduced the distance uncertainties to nearby galaxies.  This now allows detailed comparisons between observationally derived parameters for main sequence O stars and model atmosphere predictions.  Further studies incorporating improved model atmosphere flux predictions will lead to more accurate extinction curves, and positions for O stars in the log (L$_{*}$) - log(T$_{\rm eff}$) plane.  We fully expect that future detailed studies of individual luminous stars in nearby galaxies will shed important new insights on the physics of O stars and massive star formation.

The authors wish to thank Nicholas Collins, and Don Lindler for helpful suggestions about the reduction of STIS data. We acknowledge the use of the computational facilities of the Laboratory for Astronomy and Solar Physics at NASA/Goddard Space Flight Center for partial reduction and analysis of the data discussed here. We further thank our referee, Dr. Laurent Drissen, for comments and suggestions that greatly improved the manuscript. We also acknowledge support through NASA grant NAG5-3378 to Catholic University of America.






\clearpage
\begin{figure}
\epsscale{0.95}
\plotone{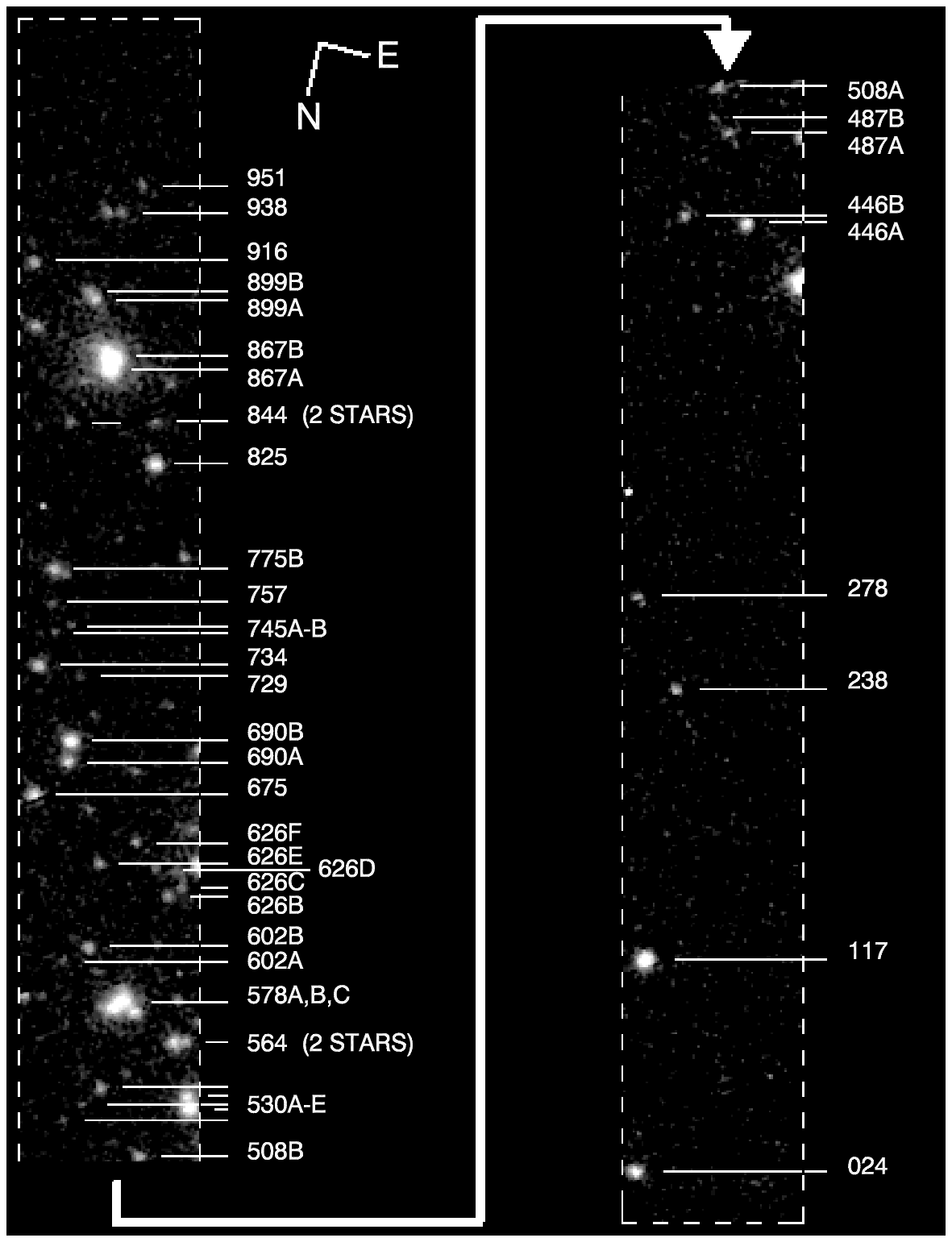}
\epsscale{1.0}
\caption{Outline of STIS aperture superposed upon 
WFPC2/F170W imagery. This figure shows the 25$\arcsec$ x 2$\arcsec$ region sampled by our STIS exposure and the approximate locations of the stars that were in the aperture corresponding to the spectral image in Figure 2.  The top of the aperture is to the left, and the bottom is to the right. The stellar objects for which we have extracted STIS/G140L spectra are identified. The names correspond to the y-position in the spectral image shown in Figure 2. The multiple alphabetical extensions to the names indicate that there are more than one source near that y-position. \label{fig1}}
\end{figure}

\clearpage
\begin{figure}
\plotone{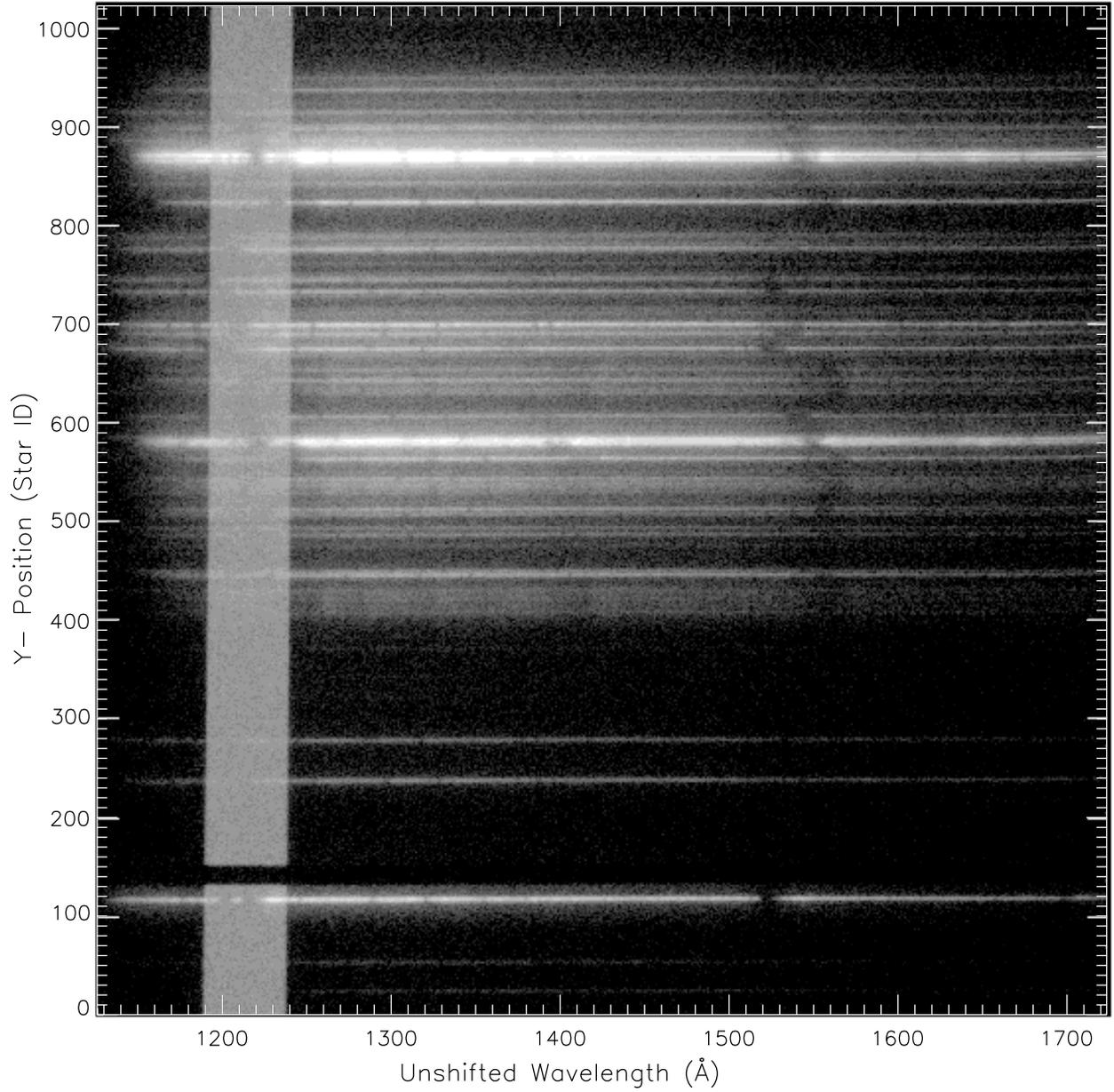}
\caption{STIS spectral image of bright O \& B stars of 
NGC~604. The figure shows the two-dimensional spectral image of NGC~604 obtained by STIS using the G140L grating. The y-position in pixel units is denoted along left side of the figure, while the unshifted approximate wavelength scale is indicated along the bottom. The strong H~I-Ly$\alpha$ background is centered at 1216\AA; its width corresponds to the 2 $\arcsec$ width of the aperture. Note the strong N~V and C~IV wind features. Strong interstellar lines can be discerned in the figure. See Figure 1 for the approximate location of the UV-bright stars in the aperture. \label{fig2}}
\end{figure}

\clearpage
\begin{figure}
\plotone{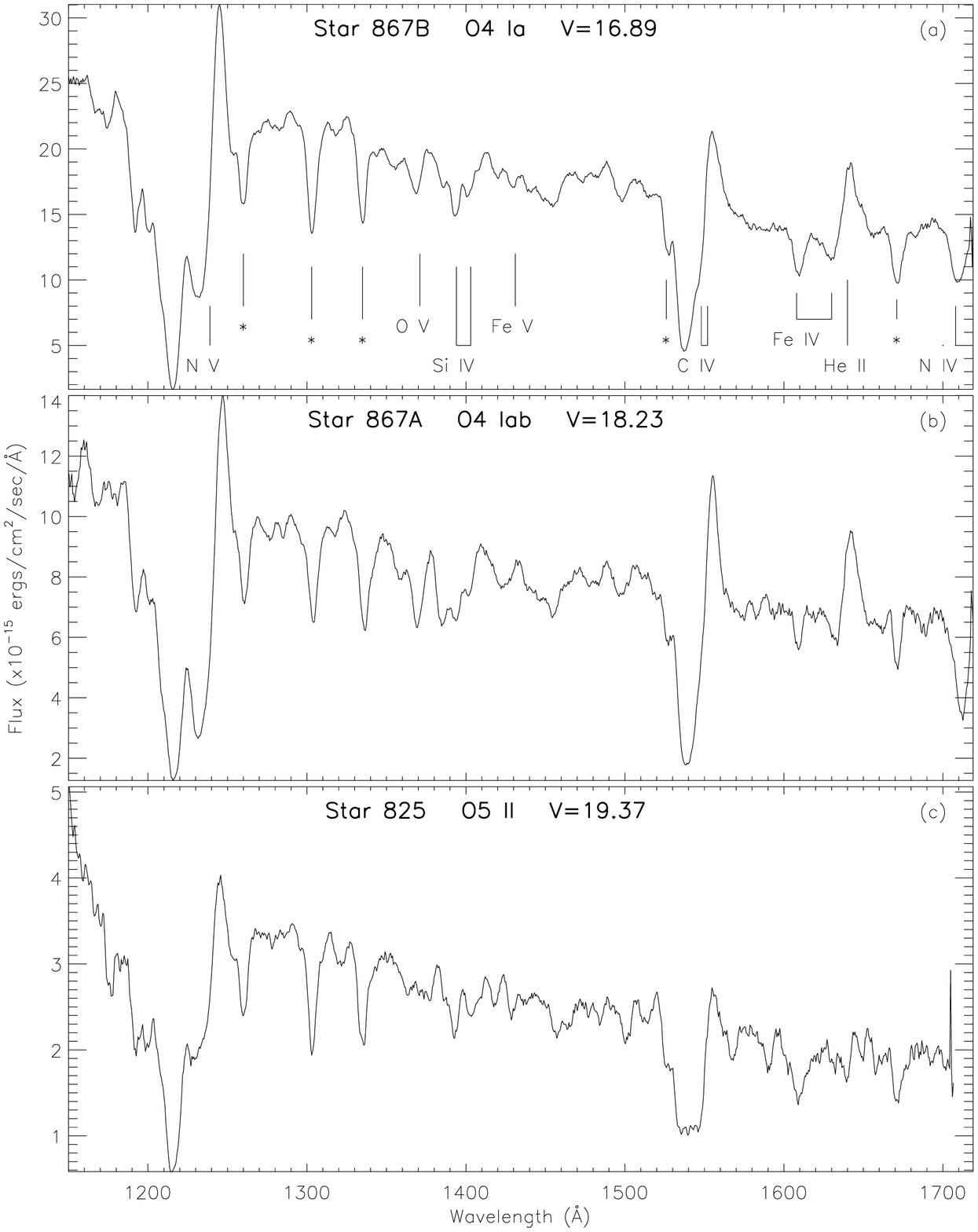}
\label{fig3a-c}
\end{figure}
\begin{figure}
\clearpage
\plotone{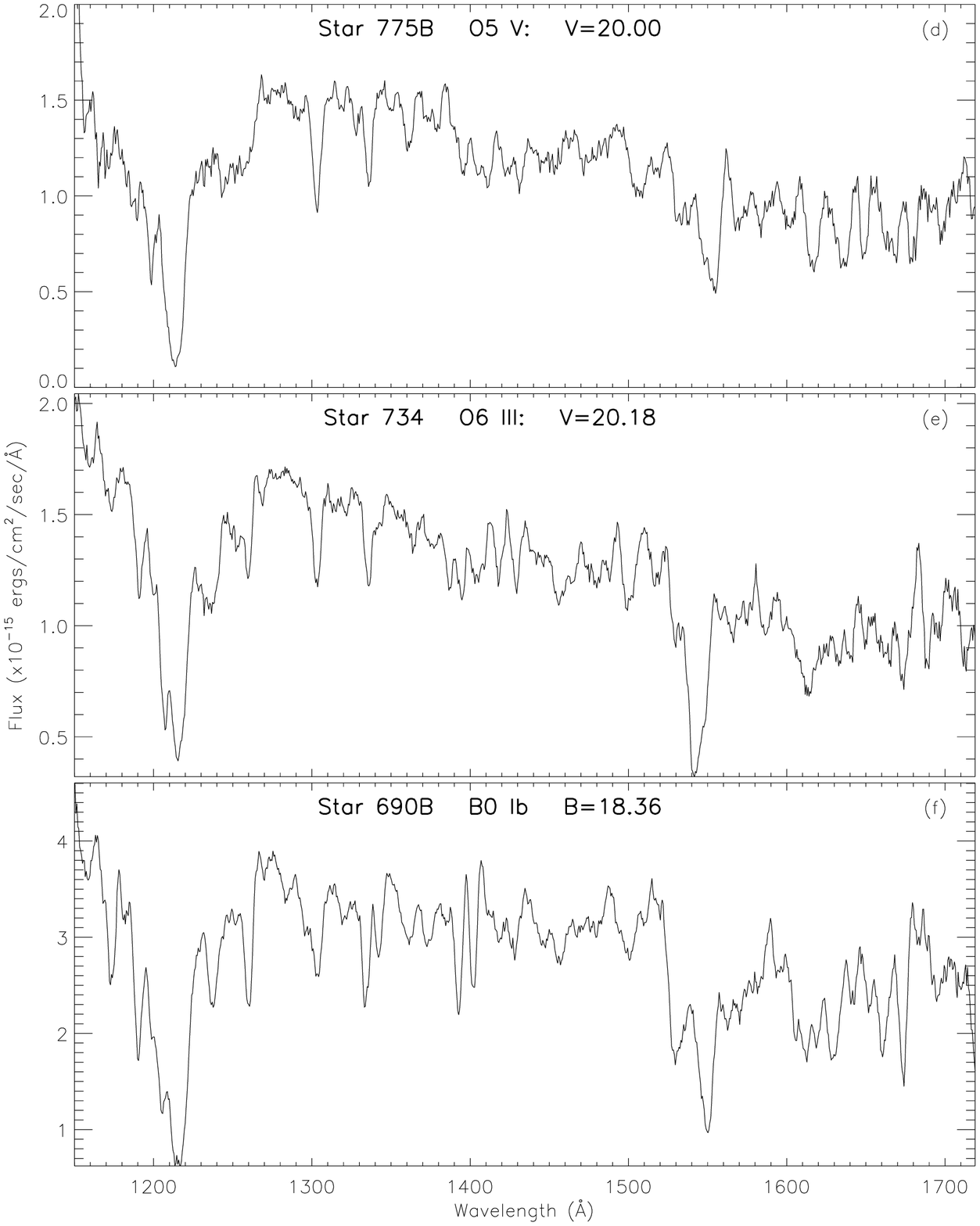}
\label{fig3d-f}
\end{figure}
\begin{figure}
\clearpage
\plotone{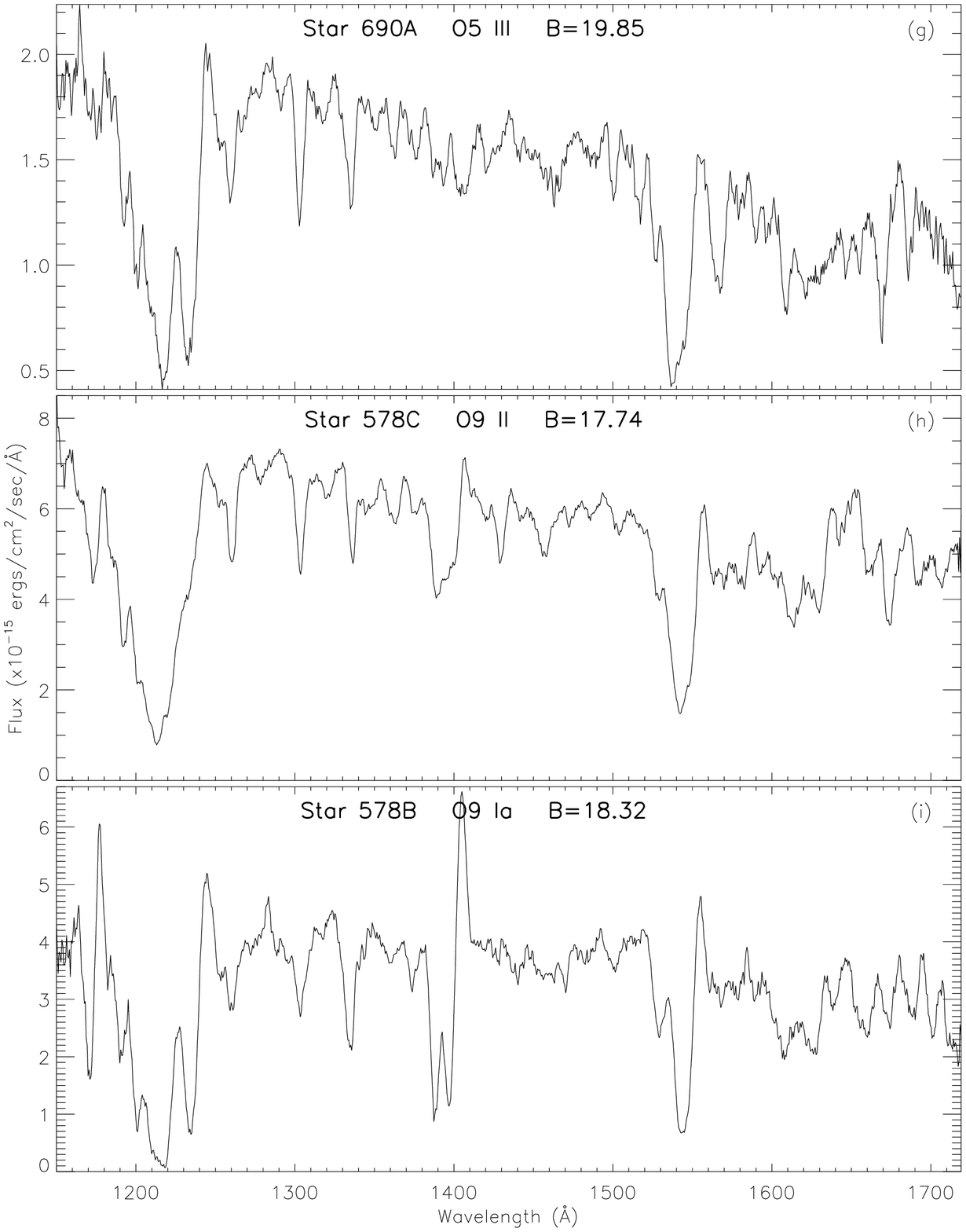}
\label{fig3g-i}
\end{figure}
\begin{figure}
\clearpage
\plotone{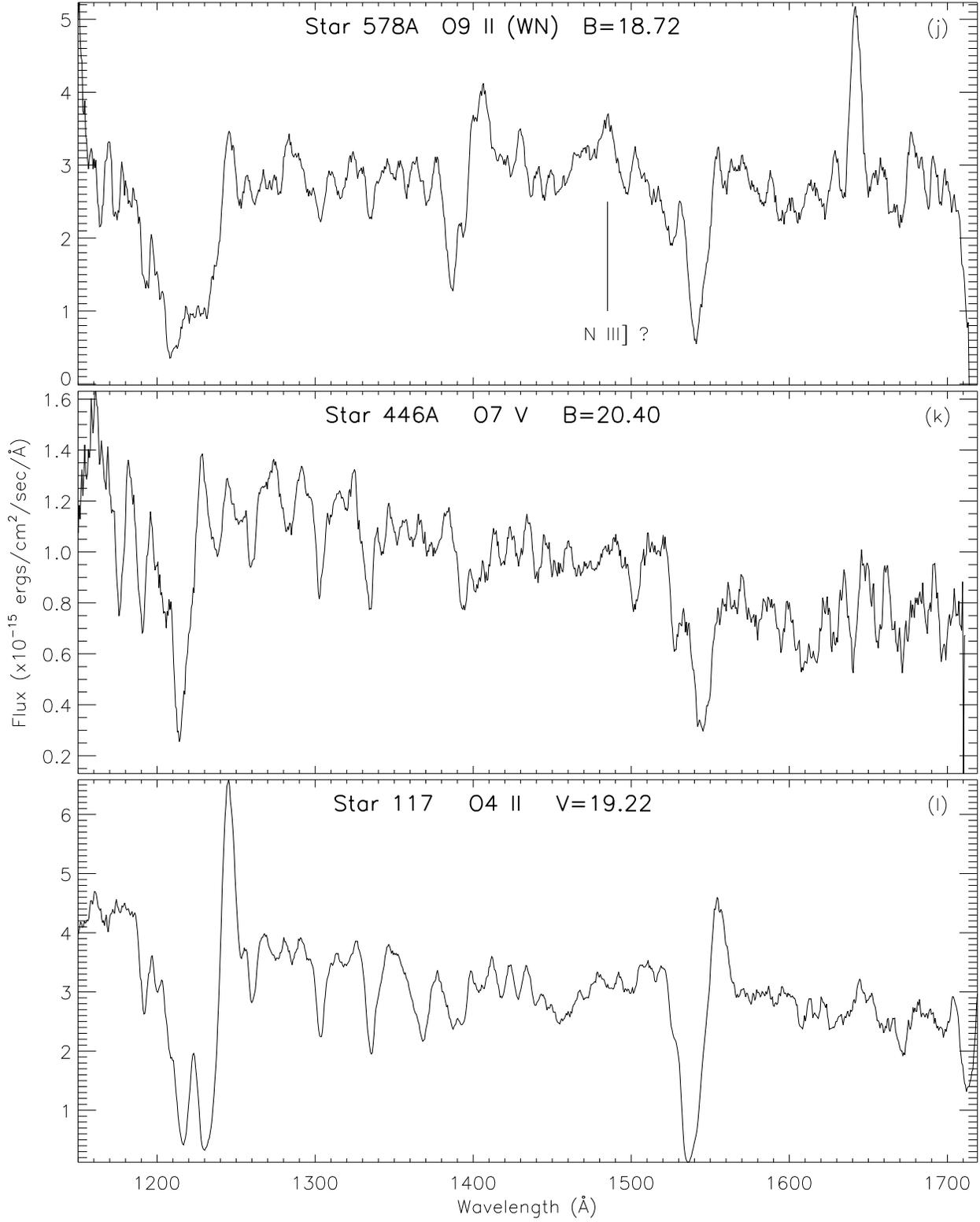}
\caption{Extracted STIS/G140L UV Spectra. In Figures 3a through 3l, we present 12 extracted UV spectra spanning the wavelength range from roughly 1150-1730\AA. In Fig 3a, we mark the positions of the strong interstellar lines that serve as wavelength fiducials for all the extracted spectra. These are denoted by asterix (*). Each panel, besides showing the extracted spectrum, gives the star ID, our deduced UV spectral type, and derived V-magnitude. \label{fig3j-l}}
\end{figure}
%
\clearpage
\begin{figure}
\hspace{-0.8in}
\includegraphics[angle=90, width=7.5in]{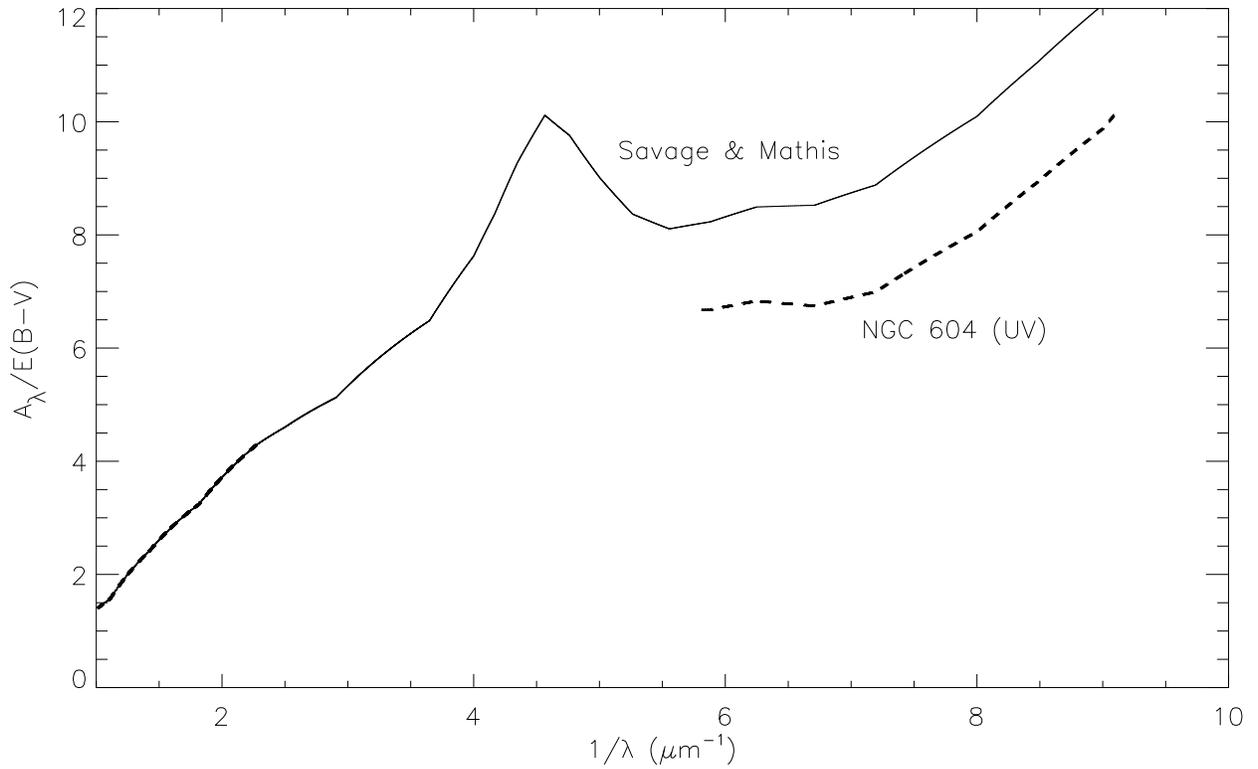}
\caption{The derived extinction curve for NGC~604 compared to that for the Milky Way. The derived extinction, A$_{\lambda}$/(E(B-V)), is plotted versus wavelength. The curve for NGC~604 is the dashed line, and the Savage \& Mathis (1979) curve is the solid line. The gap in the NGC~604 curve between 1/$\lambda$ = 2.2 to 6 ($\mu$m)$^{-1}$ indicates that we have no data sampling that wavelength region. Thus, we can say nothing about the presence of a 2175\AA\ extinction bump for NGC~604. \label{fig4}}
\end{figure}

\clearpage
\begin{figure}
\plotone{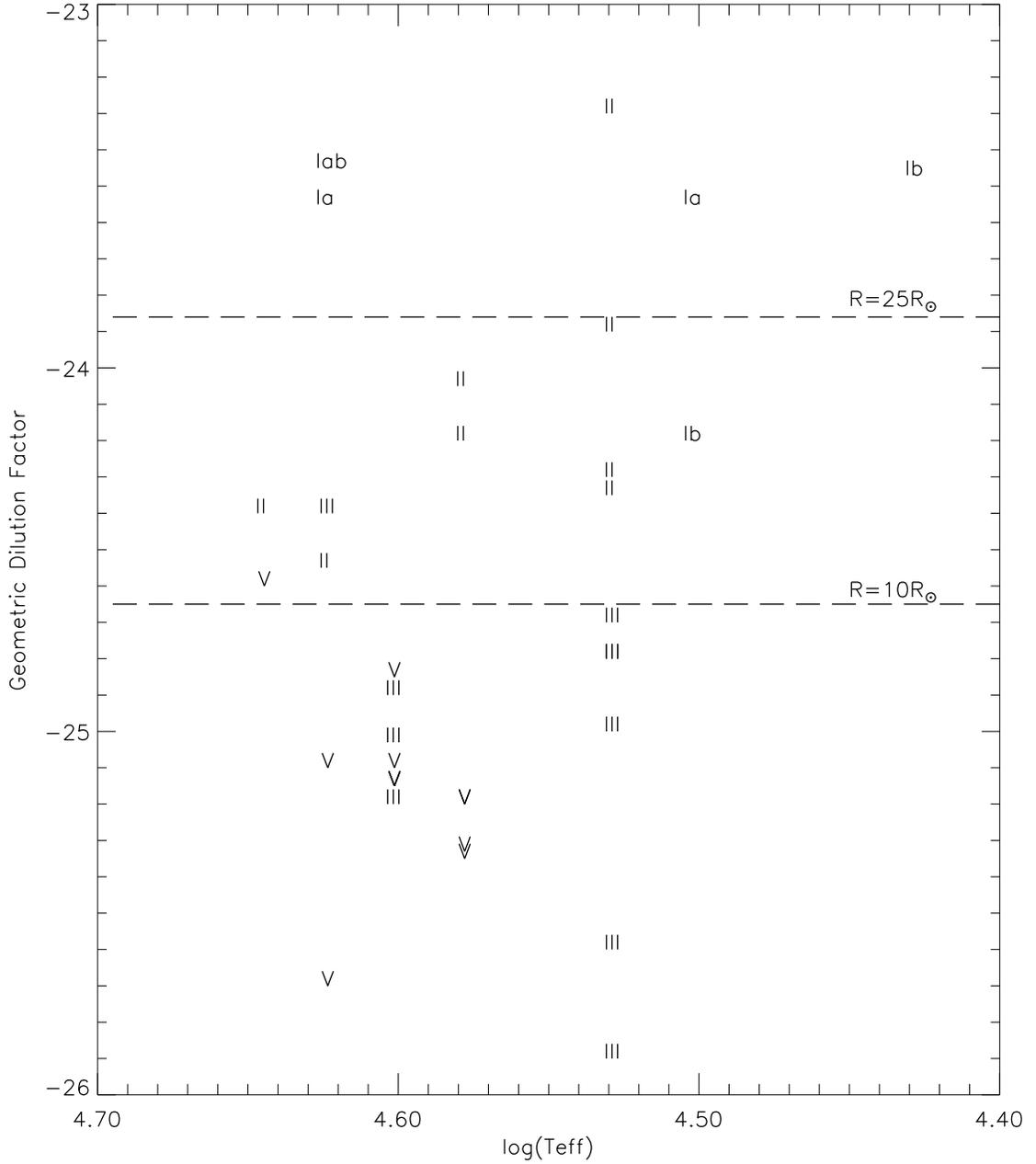}
\caption{Dilution factors versus spectral types (effective temperatures) and luminosity classes of O stars. The dilution factors derived from normalizing the dereddened, observed stellar flux distributions to the appropriate Kurucz model atmosphere are plotted for each star. The star is denoted by its luminosity class (V, III, Ia, etc.) and its position in the dilution factor (D) versus effective temperature (T$_{\rm eff}$) plane. The lines of constant stellar radius for R$_{*}$ = 10 and 25 R$_{\sun}$ are superposed on the plot. See text for discussion. \label{fig5}}
\end{figure}

\clearpage
\begin{figure}
\epsscale{0.87}
\plotone{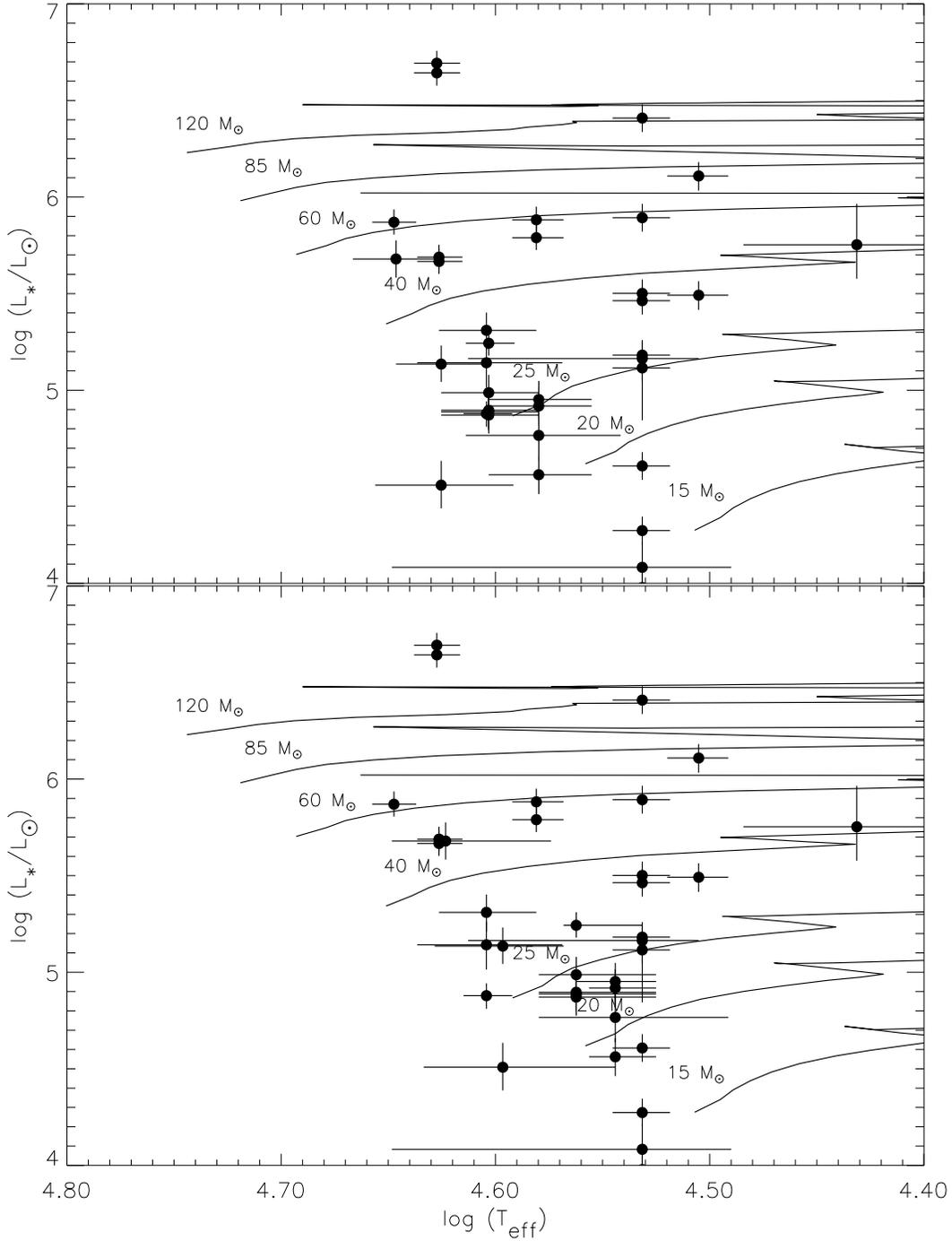}
\epsscale{1.0}
\caption{The log (L$_{*}$) versus T$_{\rm eff}$ diagram for the most luminous stars in NGC~604. The deduced luminosities and effective temperatures are given for stars with determined spectral types. The values are superposed upon the evolutionary tracks of Meynet et al. (1994). The ZAMS mass is given for each track. The leftmost extent of the tracks marks the ZAMS. In the top panel, we present the uncorrected effective temperatures and derived luminosities based upon the spectral type - effective temperature and bolometric corrections presented in Garmany \& Fitzpatrick (1990). In the lower panel, we present the revised calibrations for the main sequence O stars using the adjustments given in Martins et al. (2002). The revised calibration yields better agreement between the observationally derived parameters and the theoretical evolutionary tracks. See text for discussion. \label{fig6}}
\end{figure}
%
\clearpage
\begin{figure}
\plotone{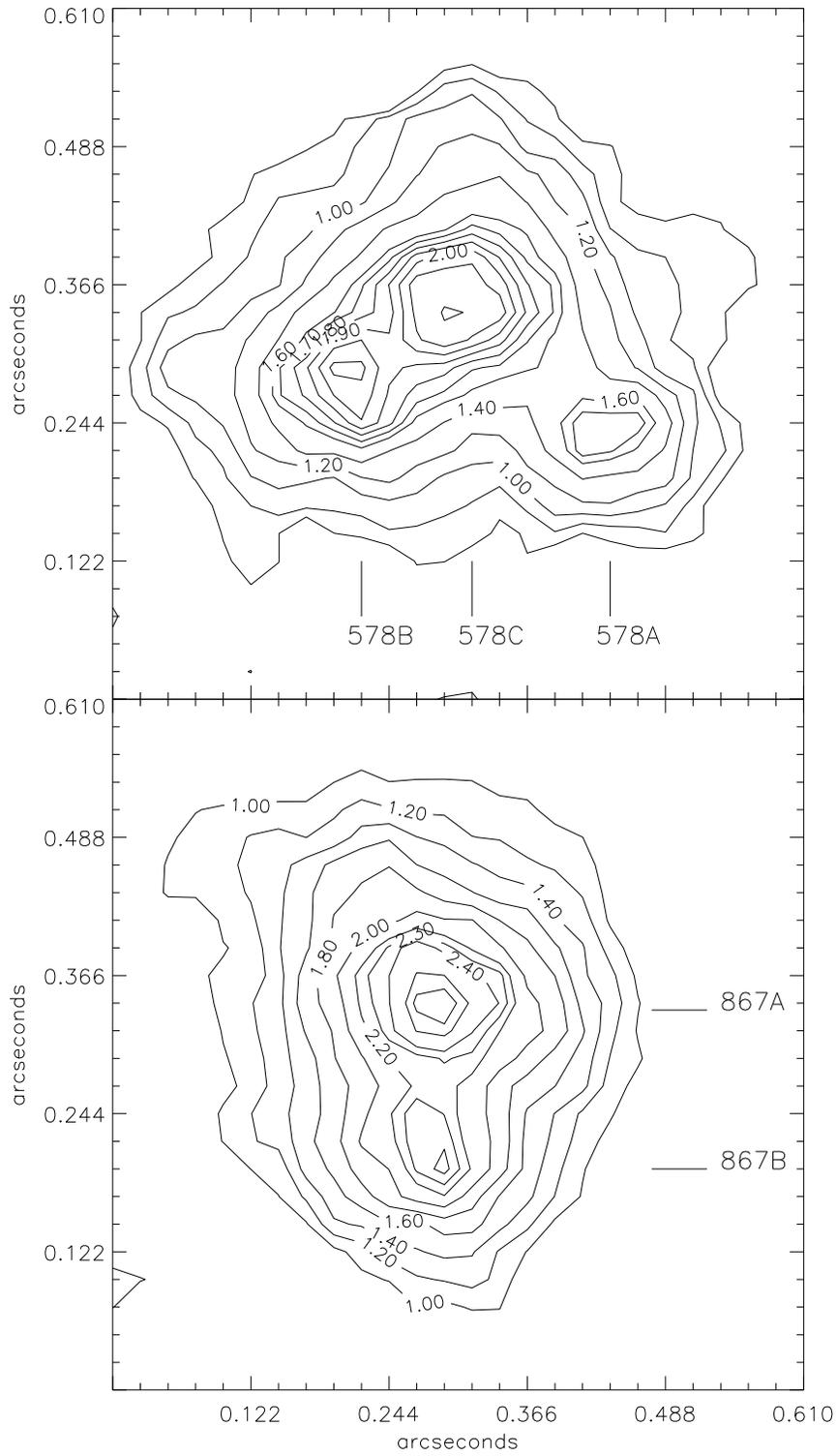}
\caption{WFPC2/F170W flux contours for the UV stellar sources, 578A, B, \& C and 867A \& 867B. The logarithmic flux contours (log (flux) -18.0) are depicted for the two groupings representing the five most luminous objects in the STIS spectral image. These data are from the PC chip. \label{fig7}}
\end{figure}

\clearpage
\begin{figure}
\plotone{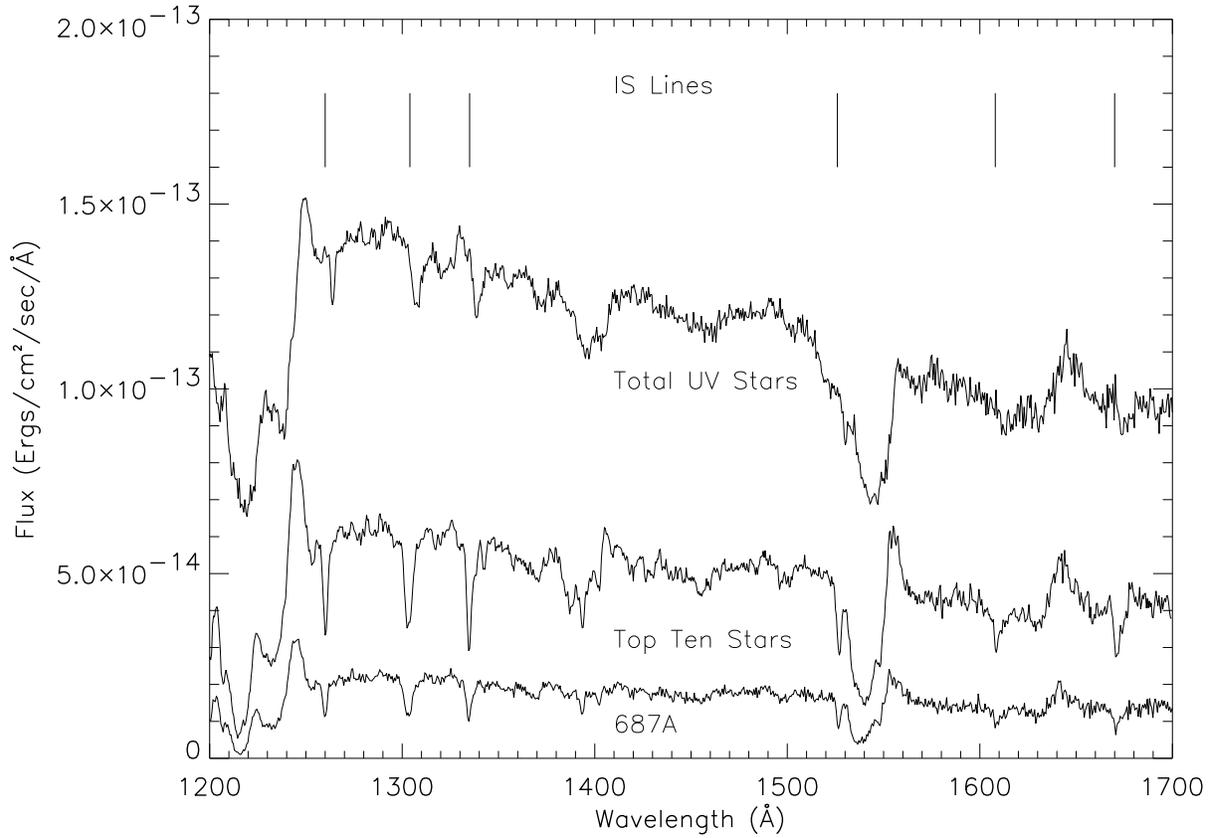}
\caption{UV spectra of brightest stars compared to total UV flux seen in the 2$\arcsec$ aperture. The UV flux of the brightest stellar object, ``867B'' is shown in the bottom spectrum. The sum of the ten brightest UV-sources (``Top 10''), identified by their 1300\AA\ fluxes in Table 2, is also plotted. The total flux in the aperture, summed perpendicular to the dispersion, is denoted. The vertical lines at the top of the figure mark positions of strong interstellar lines. \label{fig8}}
\end{figure}

\clearpage

\begin{deluxetable}{cccc}
\tablewidth{0pt}
\tablecaption{HST Datasets of NGC~604 Used}
\tablehead{\colhead{Instrument} & \colhead{Filter/Grating} & 
\colhead{Central $\lambda$} & \colhead{Dataset Name}\\  
\colhead{}& \colhead{}& \colhead{(\AA)} & \colhead{}}

\startdata
STIS &        G140L & 1425 & 	  o4x101040 \\
WFPC2 &       F555W & 5407 & 	  u2ab0201t \\
WFPC2 &       F555W & 5407 & 	  u2ab0202t \\
WFPC2 &       F814W & 7940 & 	  u2ab0203t \\
WFPC2 &       F814W & 7940 & 	  u2ab0204t \\
WFPC2 &       F656N & 6563 & 	  u2ab0205t \\
WFPC2 &       F656N & 6563 & 	  u2ab0206t \\
WFPC2 &       F336W & 3344 & 	  u2ab0207t \\
WFPC2 &       F336W & 3344 & 	  u2ab0208t \\
WFPC2 &       F170W & 1730 & 	  u2c60b01t \\
WFPC2 &       F170W & 1730 & 	  u2c60b02t \\
WFPC2 &       F547M & 5479 & 	  u2lx0307t \\
WFPC2 &       F547M & 5479 & 	  u2lx0308t \\
WFPC &        F439W & 4353 & 	  w0nn0101t \\
WFPC &        F439W & 4353 & 	  w0nn0102t \\
WFPC &        F439W & 4353 & 	  w0nn0103t \\
\enddata
\end{deluxetable}

\clearpage
\begin{deluxetable}{lcccccccc}
\tablewidth{0pt}
\tablecolumns{9}
\tablecaption{Stars Extracted from NGC~604}
\tablehead{
\colhead{Star ID} & \colhead{Drissen} & \colhead{Spectral} & \colhead{Flux} 
& 
\colhead{Flux} &
\colhead{Mag} & \colhead{B Mag} & \colhead{V Mag} & \colhead{Mag} \\
\colhead{} & \colhead{\#} & \colhead{Type} & \colhead{1300 \AA} & 
\colhead{1700 
\AA} & 
\colhead{F336W} &\colhead{F439W} & \colhead{F547M} & \colhead{F814W}}

\startdata
24&  \nodata&  B2Ia:&  7.70e-17&  1.54e-16&  \nodata&  \nodata&  \nodata&  
\nodata\\
53&  \nodata&  O9III::&  9.53E-17&  7.35E-17&  \nodata&  \nodata&  \nodata&  
\nodata\\
117&  \nodata&  04II&  3.90E-15&  2.78E-15&  \nodata&  \nodata&  19.22&  
\nodata\\
238&  \nodata&  O9III:&  5.45E-16&  5.34E-16&  \nodata&  \nodata&  20.54&  
\nodata\\
278&  \nodata&  B2Ia&  3.92E-16&  2.57E-16&  \nodata&  \nodata&  21.31&  
\nodata\\
446A&  239&  O7V&  1.22E-15&  7.71E-16&  \nodata&  20.4&  20.50&  \nodata\\
446B&  237&  O8V:&  6.62E-16&  4.44E-16&  \nodata&  21.33&  21.00&  
\nodata\\
487A&  \nodata&  O9III&  6.03E-16&  3.91E-16&  \nodata&  \nodata&  21.47&  
\nodata\\
487B&  \nodata&  O8V:&  5.82E-16&  3.97E-16&  \nodata&  \nodata&  21.89&  
\nodata\\
508A&  \nodata&  O7V:&  7.63E-16&  4.92E-16&  \nodata&  \nodata&  21.21&  
\nodata\\
508B&  227&  O6V:&  1.20E-15&  7.51E-16&  \nodata&  20.95&  20.67&  
\nodata\\
530A&  \nodata&  \nodata&  6.02E-16&  3.15E-16&  \nodata&  \nodata&  21.99&  
\nodata\\
530B&  217&  \nodata&  4.03E-16&  3.50E-16&  16.36\tablenotemark{b}&  18.62&  
18.16\tablenotemark{b}&  18.32\tablenotemark{b}\\
530C&  \nodata&  \nodata&  7.32E-16&  3.88E-16&  \nodata&  \nodata&  21.96&  
\nodata\\
530D&  216&  O7V:&  7.20E-16&  3.97E-16&  16.36\tablenotemark{b}&  19.55&  
18.16\tablenotemark{b}&  18.32\tablenotemark{b}\\
530E&  212&  \nodata&  1.31E-15&  6.07E-16&  \nodata&  20.55&  20.76&  
\nodata\\
564 (1 of 2)&  211&  O9II\tablenotemark{a}&  2.69E-15&  1.74E-15&  
17.11\tablenotemark{a}&  19.25&  18.74\tablenotemark{a}&  
18.59\tablenotemark{a}\\*
~~~~~~(2 of 2)&  210&  \nodata&  \nodata&  \nodata&  \nodata&  19.91&  
\nodata&  
\nodata\\
578A&  202&  O9II&  3.03E-15&  2.85E-15&  \nodata&  18.72&  \nodata&  
\nodata\\
578B&  200&  O9Ia&  4.48E-15&  3.30E-15&  \nodata&  18.32&  \nodata&  
\nodata\\
578C&  196&  O9II&  6.73E-15&  5.12E-15&  \nodata&  17.74&  \nodata&  
\nodata\\
602A&  191&  O8V::&  8.79E-16&  7.00E-16&  \nodata&  21.59&  21.30&  
\nodata\\
602B&  189&  O9II&  1.34E-15&  1.06E-15&  \nodata&  19.83&  19.79&  
\nodata\\
626B&  183&  O9Ib&  5.91E-16&  4.04E-16&  18.50&  20.30&  19.86&  19.68\\
626C&  \nodata&  O7V:&  6.32E-16&  3.02E-16&  \nodata&  \nodata&  \nodata&  
\nodata\\
626D&  \nodata&  \nodata&  6.62E-16&  4.34E-16&  19.94&  \nodata&  21.60&  
\nodata\\
626E&  \nodata&  \nodata&  1.12E-15&  6.09E-16&  \nodata&  \nodata&  21.03&  
\nodata\\
626F&  \nodata&  O6III&  1.03E-15&  5.82E-16&  \nodata&  \nodata&  21.26&  
\nodata\\
675&  170&  O7II&  2.87E-15&  1.87E-15&  \nodata&  19.11&  19.06&  
\nodata\\
690A&  164&  O5III&  1.89E-15&  1.19E-15&  \nodata&  19.84&  \nodata&  
\nodata\\
690B&  161&  B0Ib&  3.57E-15&  2.74E-15&  \nodata&  18.36&  \nodata&  
\nodata\\
729&  \nodata&  B2Ia:&  5.04E-16&  2.83E-16&  \nodata&  \nodata&  22.08&  
\nodata\\
734&  152&  O6III:&  1.50E-15&  9.43E-16&  \nodata&  20.36&  20.18&  
\nodata\\
745A&  \nodata&  O9III&  8.47E-16&  5.25E-16&  \nodata&  \nodata&  22.15&  
\nodata\\
745B&  \nodata&  O6V::&  8.22E-16&  4.42E-16&  \nodata&  \nodata&  22.16&  
\nodata\\
757A&  \nodata&  O7V:&  3.30E-16&  1.73E-16&  \nodata&  \nodata&  21.73&  
\nodata\\
757B&  \nodata&  O9III::&  4.83E-16&  2.84E-16&  \nodata&  \nodata&  
\nodata&  
\nodata\\
775B&  128&  O5V:&  1.56E-15&  9.64E-16&  \nodata&  20.11&  20.00&  
\nodata\\
825&  105&  O5II&  3.42E-15&  2.03E-15&  17.51&  19.26&  19.43&  19.60\\
844 (1 of 2)&  \nodata&  O8V:\tablenotemark{a}&  1.58E-15&  1.06E-15&  
19.82&  \nodata&  21.70&  22.01\\*
~~~~~~(2 of 2)&  93&  \nodata&  \nodata&  \nodata&  \nodata&  20.56&  
21.28&  \nodata\\
867A&  84&  O4IaB&  1.01E-14&  6.90E-15&  \nodata&  17.65&  \nodata&  
16.94\tablenotemark{c}\\
867B&  81&  O4Ia&  2.24E-14&  1.42E-14&  \nodata&  17.31&  \nodata&  
16.94\tablenotemark{c}\\
899A&  69&  O7II&  1.97E-15&  1.55E-15&  \nodata&  19.46\tablenotemark{d}&  
19.30\tablenotemark{d}&  \nodata\\
899B&  \nodata&  O9III&  1.03E-15&  7.51E-16&  \nodata&  \nodata&  \nodata&  
\nodata\\
916&  62&  O9III&  1.29E-15&  8.03E-16&  \nodata&  20.45&  20.20&  
\nodata\\
938 (1 of 2)&  56&  O6III::\tablenotemark{a}&  1.13E-15&  7.84E-16&  
18.78\tablenotemark{a}&  19.39&  20.67\tablenotemark{a}&  
20.00\tablenotemark{a}\\
~~~~~~(2 of 2)&  57&  \nodata&  \nodata&  \nodata&  \nodata&  20.87&  
\nodata&  \nodata\\
951&  \nodata&  O8V:&  6.94E-16&  4.40E-16&  19.34&  21.2&  21.00&  20.25\\
\enddata
\tablenotetext{a}{STIS Spectrum composed of two stars.}
\tablenotetext{b}{Magnitudes for F336W, F547M, F814W are a combination of 
two 
stars - 530B and 530D.}
\tablenotetext{c}{Magnitude for F814W is a combination of two stars - 867A 
and 
867B.}

\tablecomments{Our spectra of stars 564, 844, and 938 are composed of two 
stars 
located
horizontally next to each other on our STIS image.}
\tablecomments{The spectral types and flux values were determined from our 
extracted spectra.}
\tablecomments{The B magnitudes are from Drissen et al.  All other 
magnitudes 
are calculated from previous HST/WFPC2 imagery using HSTPhot.  The WFPC2 
filters 
used are listed.}
\end{deluxetable}
\clearpage

\begin{deluxetable}{ccccc}
\tablewidth{0pt}
\tablecolumns{5}
\tablecaption{Deduced Reddening, Effective Temperature, and Luminosity Values}
\tablehead{\colhead{Star ID}&    \colhead{log(T$\rm_{eff}$)}&   
\colhead{log(L$_{*}$/L$_{\odot}$)}&    \colhead{E(B-V)}&    \colhead{Corrected log(T$\rm_{eff}$)}}
\startdata

867A&	4.6274&      6.693&	 0.30&  	\nodata\\
867B&	4.6274&      6.643&	 0.15&  	\nodata \\
578C&	4.5315&      6.409&	 0.30&   	\nodata\\
578B&	4.5052&      6.109&	 0.30&   	\nodata\\
578A&	4.5315&      5.893&	 0.20&   	\nodata\\
675& 	4.5809&      5.882&	 0.20&   	\nodata\\
117& 	4.6474&      5.870&	 0.10&   	\nodata\\
899A&	4.5809&      5.782&	 0.20&   	\nodata\\
690B&	4.4314&      5.753&	 0.20&   	\nodata\\
690A&	4.6263&      5.689&	 0.20&   	\nodata\\
775B&	4.6464&      5.679&	 0.20&         4.6233\\
825& 	4.6263&      5.667&	 0.10&   	\nodata\\
564& 	4.5315&      5.493&	 0.10&   	\nodata\\
626B&	4.5052&      5.492&	 0.30&   	\nodata\\
602B&	4.5315&      5.463&	 0.20&   	\nodata\\
734& 	4.6042&      5.310&	 0.05&   	\nodata\\
446A&	4.6031&      5.243&	 0.10&         4.5623\\
899B&	4.5315&      5.174&	 0.05&   	\nodata\\
238& 	4.5315&      5.164&	 0.21&   	\nodata\\
508B&	4.6253&      5.135&	 0.025& 	4.5966\\
938& 	4.6042&      5.134&	 0.10&   	\nodata\\
916& 	4.5315&      5.115&	 0.05&   	\nodata\\
530D&	4.6031&      4.979&	 0.05&         4.5623\\
446B&	4.5798&      4.952&	 0.075& 	4.5441\\
951& 	4.5798&      4.918&	 0.05&         4.5441\\
508A&	4.6031&      4.896&	 0.05&         4.5623\\
626C&	4.6031&      4.880&	 0.05&         4.5623\\
626F&	4.6042&      4.879&	 0.05&   	\nodata\\
757A&	4.6031&      4.871&	 0.20&         4.5622\\
602A&	4.5798&      4.766&	 0.025& 	4.5441\\
487A&	4.5315&      4.608&	 0.05&   	\nodata\\
487B&	4.5798&      4.562&	 0.05&         4.5441\\
745B&	4.6253&      4.508&	 0.0&	       4.5966\\
745A&	4.5315&      4.273&	 0.0&   	\nodata\\
53&  	4.5315&      4.075&	 0.05&   	\nodata\\
278& 	4.2305&      4.010&	 0.1&   	\nodata\\
757B&	4.5315&      3.970&	 0.0&   	\nodata\\
24&  	4.2305&      3.636&	 0.3&	 	\nodata\\*
729& 	4.2305&      3.576&	 0.0&   	\nodata\\*
\enddata
\end{deluxetable}
\clearpage
\begin{deluxetable}{cccc}
\tablewidth{0pt}
\tablecolumns{4}
\tablecaption{NGC~604 Stars versus Mass Range \& IMF Predictions}
\tablehead{\colhead{Mass Range} & \colhead{Miller \& Scalo} & \colhead{Salpeter IMF} & 
\colhead{Observations} \\
\colhead{(M$_{\sun}$)} & \colhead{$\alpha$ = -2.7} & \colhead{$\alpha$ = -2.3} & \colhead{}}
\startdata
20-25 	&	9.5 &	7.6 &	7 \\
25-40 	&	11.3 & 10.3 & 11 \\
40-60 	&	4.6 &	5.0 &	7 \\
60-85 	&	2.1 &	2.6 &	2 \\
 85-120 &	1.1 &	1.7 &	0 \\
120-150 &	1.4 &	2.9 &	3 \\
\enddata
\end{deluxetable}

\end{document}